\documentclass[showpacs,prl,onecolumn,aps,superscriptaddress,preprintnumbers,nofootinbib]{revtex4}
\usepackage[T1]{fontenc}
\usepackage[latin9]{inputenc}
\usepackage{amsmath,amssymb}
\usepackage{epsfig}
\usepackage{graphicx}
\usepackage{amsmath}
\usepackage{amsfonts}
\def\slashchar#1{\setbox0=\hbox{$#1$}     		% set a box for #1
   \dimen0=\wd0                                 	% and get its size
   \setbox1=\hbox{/} \dimen1=\wd1               	% get size of /
   \ifdim\dimen0>\dimen1                        	% #1 is bigger
      \rlap{\hbox to \dimen0{\hfil/\hfil}}      	% so center / in box
      #1                                        	% and print #1
   \else                                        	% / is bigger
      \rlap{\hbox to \dimen1{\hfil$#1$\hfil}}   	% so center #1
      /                                         	% and print /
   \fi}

\renewcommand{\vec}{\boldsymbol}
\newcommand{\beq}{\begin{equation}}
\newcommand{\eeq}{\end{equation}}
\newcommand{\bea}{\begin{eqnarray}}
\newcommand{\eea}{\end{eqnarray}}
\newcommand{\ba}{\begin{array}}
\newcommand{\ea}{\end{array}}

\def\eq#1{{Eq.~(\ref{#1})}}
\def\fig#1{{Fig.~\ref{#1}}}
\newcommand{\bas}{\bar{\alpha}_S}
\newcommand{\as}{\alpha_S}
\newcommand{\nn}{\nonumber}

\newcommand{\h}{\frac{1}{2}}

\newcommand{\ga}{\gamma}

\newcommand{\Lb}{\left(}
\newcommand{\Rb}{\right)}
\def\pom{{I\!\!P}}

\begin{document}

\title{Perturbative QCD and beyond: azimuthal angle correlations in 
 deuteron-deuteron scattering   from Bose-Einstein correlations.}
\author{E. ~Gotsman}
\email{gotsman@post.tau.ac.il}
\affiliation{Department of Particle Physics, School of Physics and Astronomy,
Raymond and Beverly Sackler
 Faculty of Exact Science, Tel Aviv University, Tel Aviv, 69978, Israel}
 \author{ E.~ Levin}
\email{leving@post.tau.ac.il, eugeny.levin@usm.cl}
\affiliation{Department of Particle Physics, School of Physics and Astronomy,
Raymond and Beverly Sackler
 Faculty of Exact Science, Tel Aviv University, Tel Aviv, 69978, Israel}
 \affiliation{Departemento de F\'isica, Universidad T\'ecnica Federico
 Santa Mar\'ia, and Centro Cient\'ifico-\\
Tecnol\'ogico de Valpara\'iso, Avda. Espana 1680, Casilla 110-V,
 Valpara\'iso, Chile} 
 
\date{\today}


\begin{abstract}
In this paper, we found within the framework of perturbative QCD, that
 in deuteron-deuteron scattering 
 the Bose-Einstein correlations due to two parton showers production, 
induce  azimuthal angle correlations, with three correlation lengths:
 the size of the deuteron ($R_D$), the proton radius ($R_N$), and the
 size of the BFKL
 Pomeron which, is  closely   related to the saturation momentum
 ($R_c \sim 1/Q_s$). These correlations are independent of the values
 of rapidities of the produced gluons (long range rapidity correlations), 
for
 large rapidities ($\bas |y_1 - y_2| \geq 1$),
 and have no symmetry with respect to $\phi \to  \pi  -
 \phi$ ($ \vec{p}_{T1}  \to  - \vec{p}_{T1}$).  Therefore,
 they give rise to $v_n$ for all values of $n$,  not only
 even values.  The contributions with the
 correlation length $R_D$ and $R_N$
 crucially depend on the non-perturbative contributions,  and to  
obtain estimates of their values, 
requiries a lot of modeling, while the correlations with  $R_c \sim 
1/Q_s$
 have a perturbative QCD origin, and can be  estimated in the Color
 Glass Condensate (CGC) approach.
 
  \end{abstract}
 
 \preprint{TAUP-3007/16}

\maketitle

%\tableofcontents

%%%%%%%%%%%%%%%%%%%%%%%%%%%%%%%%%%%%%%%%%%%%%%%
\section{Introduction}
%%%%%%%%%%%%%%%%%%%%%%%%%%%%%%%%%%%%%%%%%%%%%%%
In this paper we continue to resurrect the old ideas of Gribov Pomeron
 Calculus, that the Bose
-Einstein correlations lead to strong azimuthal angle
 correlations\cite{PION}, which do not depend on the rapidity
 difference between measured hadrons ( large range rapidity
 (LRR) correlations).  In the framework of QCD, these azimuthal
 correlations stem from the production of two patron showers,
 and have been re-discovered in Refs.\cite{KOWE,KOLUCOR}. In
 Ref.\cite{GLMBE} it  was demonstrated 
that  Bose--Einstein correlations  generate $v_n$ with even and odd $n$,
 with values which are close to the experimental values 
 \cite{CMSPP,STARAA,PHOBOSAA,STARAA1,CMSPA,CMSAA,ALICEAA,ALICEPA,
ATLASPP,ATLASPA,ATLASAA}.

The goal of this paper is to show that the  Bose--Einstein correlations
 that have been discussed in Refs.\cite{PION,GLMBE},  arise naturally  
in the
 perturbative QCD approach, together with ones that have been considered
 in Refs.\cite{KOWE,KOLUCOR}. We believe that the qualitative difference
 between these two approaches  originates   from different sources of 
the
 Bose-Einstein correlations: the two parton shower production in
 Refs.\cite{PION,GLMBE} and  one parton shower for  Refs.\cite{KOWE,KOLUCOR}.

  We consider  here the azimuthal correlations for   
deuteron-deuteron
 scattering at high energy. It is well known\cite{HBT}, (see also Refs.
\cite{IPCOR}) that Bose-Einstein correlations  provide a possibility 
to
 measure the volume of interaction or, in other words, the typical
 sizes of the interaction. Indeed,  the general formula for the Bose-Einstein
 correlations \cite{HBT,IPCOR} takes the form
 \beq \label{I1}
  \frac{d^2 \sigma}{d y_1 \,d y_2  d^2 p_{T1} d^2 p_{T2}}\Lb \rm identical\,\,
 gluons\Rb\,\,\propto\,\,\Big{ \langle}  1\,\,+\,\,e^{i
 r_\mu Q_\mu}\Big{\rangle}
 \eeq
 where  averaging $\langle \dots \rangle$ includes the integration
 over $r_\mu = r_{1,\mu} - r_{2,\mu}$.  For the case of $y_1=y_2$,  
  $Q_\mu = p_{1,\mu} \,-\,p_{2,\mu}$ simplifies to $\vec{Q}\,\equiv
 \,\vec{p}_{T,12}\,=\,\vec{p}_{T1} \,-\,\vec{p}_{T2}$,
 
 One can see that \eq{I1} allows us to measure the typical $r_\mu$
 for the interaction. For deuteron-deuteron scattering we expect
 several typical $r$: the size of the deuteron $R_D$, the nucleon
 size  $R_N$, and the typical size, related to the saturation scale
 ($r_{\rm sat}\,=1/Q_s$, where $Q_s$  denotes the saturation 
scale\cite{KOLEB}).
 In our calculation we   hope to see  the appearance of these 
scales.
 
\begin{figure}[ht]
 \includegraphics[width=14cm]{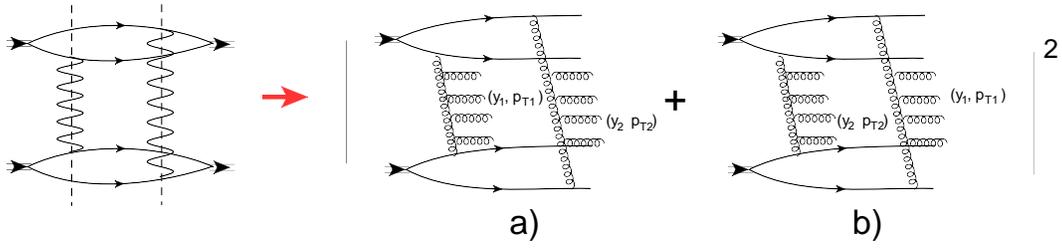}
   \protect\caption{ The two parton showers production that contributes
 to the Glauber correction term for deuteron-deuteron scattering. The
  wavy lines describe the exchange of the BFKL Pomeron. \fig{ddagk}-a
 and \fig{ddagk}-b show two diagrams that can interfer  for 
 identical gluons. The dashed lines show the cut BFKL  Pomeron
\cite{AGK}.}
\label{ddagk}
   \end{figure}

It is well known, that the total cross section for the deuteron-deuteron
 scattering can be written in the form: $\sigma_{DD}\,=\,4 \sigma_{NN}
 \,-\,\Delta \sigma_{DD}$, where $\Delta \sigma_{DD}$ is the Glauber
 correction term\cite{GLA} which is proportional to $1/R^2_D$, while
 $\sigma_{NN}$  denotes the total cross section of the 
nucleon-nucleon interaction.
   Intuition,  suggests that the correlation radius of the order of 
$R_D$,
 stems from the production due to the Glauber correction term (see \fig{ddagk})
%%%%%%%%%%%%%%%%%%%%%%%%%%%%%%%%%%%%%%%%%%%%%%%%%%%%%%

 %%%%%%%%%%%%%%%%%%%%%%%%%%%%%%%%%%%%%%%%%%%%%%%%
 
 The production of two gluon are shown in \fig{ddagk}-a and \fig{ddagk}-b,
where interference in the case of the generated identical gluon 
 leads to
 the correlation function of \eq{I1}. Generally speaking, the  
inclusive
 production of two gluons with rapidities $y_1$ and $y_2$ and transverse
 momenta $\vec{p}_{T1}$ and $\vec{p}_{T2}$, takes the form
 \bea \label{I2}
 &&  \frac{d^2 \sigma}{d y_1 \,d y_2  d^2 p_{T1} d^2 p_{T2}}\Lb \rm
 identical\,\,
 gluons\Rb\,\,=\\
 &&  \frac{d^2 \sigma}{d y_1 \,d y_2  d^2 p_{T1} d^2 p_{T2}}\Lb \rm
 different \,\,
 gluons\Rb \,\Bigg( \underbrace{1}_{\rm squared\,  of\,
 diagrams}\,\,\,+\,\,\underbrace{C\Lb R_c |\vec{p}_{T1}\,-\,\vec{p}_{T2}|
\Rb}_{\rm interference\,diagram}\Bigg)\nn
 \eea
In \eq{I2}  $R_c$  denotes the correlation radius (correlation length), 
and in
 the form of the correlation function, we anticipate that the production
 of two parton showers leads to the double inclusive cross section, that
 does not depend on rapidities $y_1$ and $y_2$.
  
%%%%%%%%%%%%%%%%%%%%%%%%%%%%%%%%%%%%%%%%%%%%%%%
\section{Born Approximation}
%%%%%%%%%%%%%%%%%%%%%%%%%%%%%%%%%%%%%%%%%%%%%%% 
 
\begin{figure}[ht]
 \includegraphics[width=16cm]{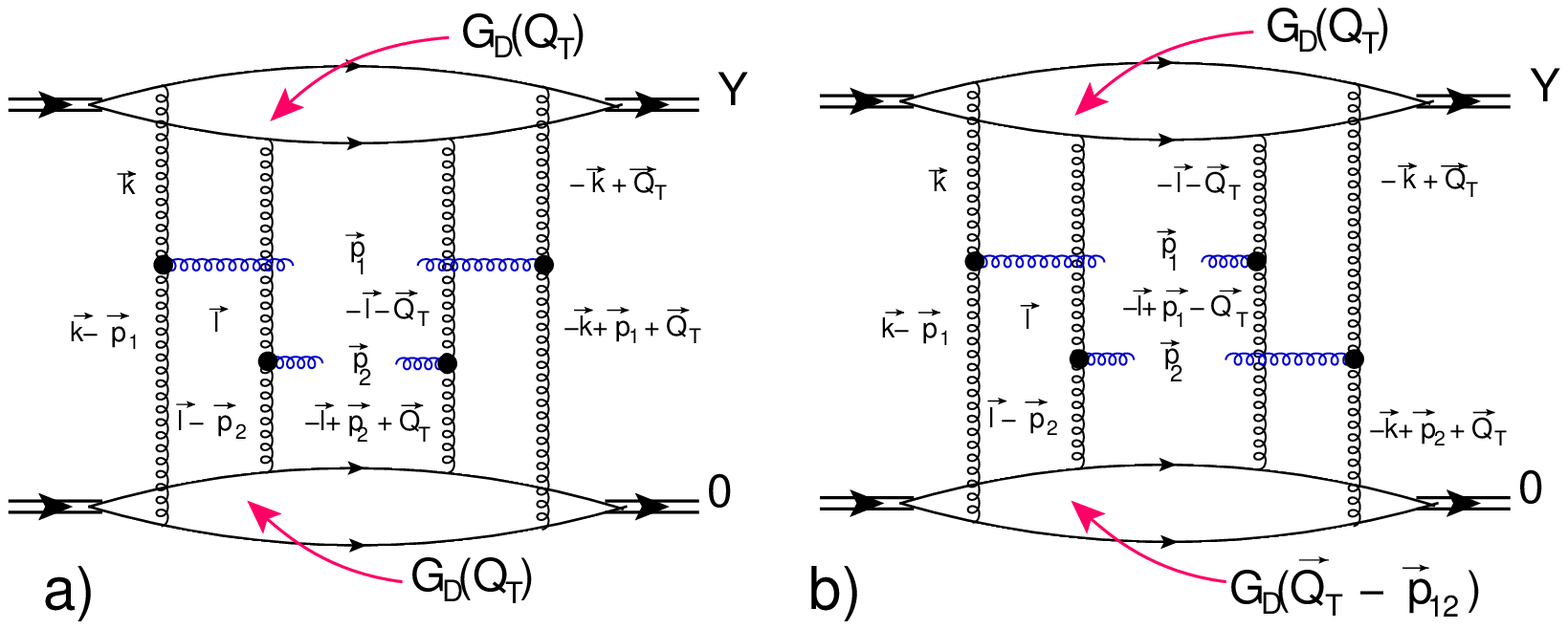}
   \protect\caption{ The double inclusive production of two gluons
 with rapidities $y_1$ and $y_2$ and transverse momenta $\vec{p}_{T1}$
 and $\vec{p}_{T2}$, in the Born Approximation of perturbative QCD. The
 interference diagram of \fig{ddba}-b gives the correlation function
 $C\Lb R_D |\vec{p}_{T1}\,-\,\vec{p}_{T2}|\Rb$ of \eq{I2}. 
 The  solid lines denote nucleons in the deuterons,
 which are  specified by double lines}
\label{ddba}
   \end{figure}

 %%%%%%%%%%%%%%%%%%%%%%%%%%%%%%%%%%%%%%%%%%%%%%%
{\boldmath{\subsection{Bose-Einstein correlation function with
 radius $\propto\,\, R_D$}}}
%%%%%%%%%%%%%%%%%%%%%%%%%%%%%%%%%%%%%%%%%%%%%%% 
 The simplest contribution in the Born approximation of perturbative
 QCD {\bf} is shown in \fig{ddba}. The second diagram describes the 
interference  between  two parton showers,  shown in \fig{ddagk}-b.

 %%%%%%%%%%%%%%%%%%%%%%%%%%%%%%%%%%%%%%%%%%%%%%%%%%%%%%

 %%%%%%%%%%%%%%%%%%%%%%%%%%%%%%%%%%%%%%%%%%%%%%%%
 
 The analytical expressions take the following forms. For the diagram
 of \fig{ddba}-a we have
 \bea \label{BAD1}
  & &\frac{d^2 \sigma}{d y_1 \,d y_2  d^2 p_{T1} d^2 p_{T2}}\Lb
 \fig{ddba}-a\Rb\,\,\propto\,\int d^2 Q_T G^2_D\Lb Q_T\Rb\nn\\
  &\times&  \int d^2k_t\,d^2 l_T\Bigg( \frac{I_P\Lb \vec{k}_T, -\vec{k}_T +
 \vec{Q}_T\Rb}{\vec{k}^2_T\,\Lb \vec{k}-\vec{p}_1\Rb^2_T}\Gamma_\mu\Lb
 \vec{k}_T, \vec{p}_{T1}\Rb\,\Gamma_\mu\Lb -\vec{k}_T +\vec{Q}_T,
 \vec{p}_{T1}\Rb\,\frac{I_P\Lb \vec{k}_T - \vec{p}_{T1}, -\vec{k}_T 
  +\vec{p}_{T1}+\vec{Q}_T\Rb}{\Lb -\vec{k}+\vec{Q}\Rb^2_T\,\Lb 
-\vec{k}+\vec{p}_{1}+\vec{Q}\Rb^2_T}\, \Bigg)\nn\\
 &\times& 
\Bigg( \frac{I_P\Lb \vec{l}_T, -\vec{l}_T - \vec{Q}_T\Rb}{\vec{l}^2_T\,\Lb
 \vec{l}-\vec{p}_2\Rb^2_T}\Gamma_\nu\Lb \vec{l}_T, \vec{p}_{T2}\Rb\,
\Gamma_\nu\Lb -\vec{l}_T -\vec{Q}_T, \vec{p}_{T2}\Rb\,\frac{I_P\Lb
 \vec{l}_T - \vec{p}_{T2}, -\vec{l}_T   +\vec{p}_{T2} - \vec{Q}_T\Rb}{\Lb
 -\vec{l}-\vec{Q}\Rb^2_T\,\Lb -\vec{l}+\vec{p}_{2}-\vec{Q}\Rb^2_T}\,
 \Bigg)\eea

The interference diagram of \fig{ddba}-b takes the following form
 \bea \label{BAD2}
  & &\frac{d^2 \sigma}{d y_1 \,d y_2  d^2 p_{T1} d^2 p_{T2}}\Lb \fig{ddba}-b\Rb\,\,\propto\,\int d^2 Q_T G_D\Lb Q_T\Rb \,G_D\Lb \vec{Q}_T\,+\,\vec{p}_{T,12}\Rb\nn\\
  &\times&  \int d^2k_T\,d^2 l_T\Bigg( \frac{I_P\Lb \vec{k}_T, -\vec{k}_T + \vec{Q}_T\Rb}{\vec{k}^2_T\,\Lb \vec{k}-\vec{p}_1\Rb^2_T}\Gamma_\mu\Lb \vec{k}_T, \vec{p}_{T1}\Rb\,\Gamma_\mu\Lb -\vec{l}_T -\vec{Q}_T, \vec{p}_{T1}\Rb\,\frac{I_P\Lb \vec{k}_T - \vec{p}_{T1}, -\vec{k}_T   +\vec{p}_{T1}+\vec{Q}_T\Rb}{\Lb -\vec{l}-\vec{Q}\Rb^2_T\,\Lb -\vec{l}+\vec{p}_{1}-\vec{Q}\Rb^2_T}\, \Bigg)\nn\\
 &\times& 
\Bigg( \frac{I_P\Lb \vec{l}_T, -\vec{l}_T - \vec{Q}_T\Rb}{\vec{l}^2_T\,\Lb \vec{k}-\vec{p}_2\Rb^2_T}\Gamma_\mu\Lb \vec{l}_T, \vec{p}_{T2}\Rb\,\Gamma_\mu\Lb -\vec{k}_T +\vec{Q}_T, \vec{p}_{T2}\Rb\,\frac{I_P\Lb \vec{l}_T - \vec{p}_{T2}, -\vec{l}_T   +\vec{p}_{T2} - \vec{Q}_T\Rb}{\Lb -\vec{k}+\vec{Q}\Rb^2_T\,\Lb -\vec{k}+\vec{p}_{2}+\vec{Q}\Rb^2_T}\, \Bigg)\eea

 The Lipatov vertices $\Gamma_\mu
 $ have the form (see  reference \cite{KOLEB} for example):

   \beq \label{LIP}
   \Gamma_\mu\Lb k_T, p_{T1}\Rb\,\,=\,\,\frac{1}{p_{T1}^2}\Lb \,k^2_T\,
\vec{p}_{T1}   \,-\,\vec{k}_T\,p^2_{T1}\Rb;~~~~~~~~
     \Gamma_\mu\Lb k_{T1},p_{T2}\Rb\,\,=\,\,\frac{1}{p_{T2}^2}\Lb \,k^2_{T1}\,
\vec{p}_{T2}   \,-\,\vec{k}_{T1}\,p^2_{T2}\Rb;
      \eeq
      
      and
      
      \beq \label{A2}
         \Gamma_\mu\Lb k_T,p_{T1}\Rb\,  \Gamma_\mu\Lb k_{T1},p_{T2}\Rb
 \,\,=\,\,\frac{k^2_{T1}\,\Lb \vec{k}_T - \vec{p}_{T2}\Rb^2}{p^2_{T2}}
     \,+\,\frac{k^2_{T}\,\Lb \vec{k}_{T1} - \vec{p}_{T1}\Rb^2}{p^2_{T1}}
  \,-\,Q^2_T\,-\,p^2_{T,12}\frac{k^2_T\,k^2_{T1}}{p^2_{T1}\,p^2_{T2}}
         \eeq
         
         where $\vec{p}_{T,12} \,=\,\vec{p}_{T1} - \vec{p}_{T2}$   
         and  $\vec{k}_{T1} = \vec{k}_T - \vec{Q}_T$. One can see  from \eq{A2} that
         \bea \label{A21}
            \Gamma_\mu\Lb k_T,p_{T1}\Rb\,  \Gamma_\mu\Lb k_{T1},p_{T2}\Rb\,\,&\xrightarrow{k_T \,\ll\,Q_T}\, &     \,\,k^2_T \Lb 1 + {\cal O}\Lb \frac{Q}{p_{T1}}; \frac{k_T}{p_{T1}}\Rb\Rb\nn\\
                  \Gamma_\mu\Lb k_T,p_{T1}\Rb\,  \Gamma_\mu\Lb k_{T1},p_{T2}\Rb\,\,&\xrightarrow{k_{T1} \,\ll\,Q_T}\,  &    \,\,k^2_{T1} \Lb 1 + {\cal O}\Lb \frac{Q}{p_{T2}}; \frac{k_{T1}}{p_{T2}}    \Rb\Rb;
            \eea

   $G_D\Lb Q_T\Rb$ is equal to
   \beq \label{BAD3}
       G_D\Lb Q_T\Rb\,\,=\,\,\int d^2 r\,e^{i \vec{r} \cdot \vec{Q}_T}|\Psi_D\Lb r\Rb|^2\,
       \eeq
       where $r$ denotes the distance between the proton and the neutron 
in the deuteron.
  The impact factors
       ($I_P\Lb \vec{k}_T, -\vec{k}_T + \vec{Q}_T\Rb $ and others in
 \eq{BAD1} and \eq{BAD2}), determine the interaction of two gluons with
 the nucleon, and their typical momenta   are about $1/R_N$.    
       
       From \eq{BAD3} we can see that typical 
       $Q_T \propto\,1/R_D$, where $R_D$ is the deuteron radius. In
 other words, $Q_T$  (and $|\vec{Q}_T \,+\,\vec{p}_{T,12}|$ in
 \eq{BAD2})    turn out to be much smaller than the value of
 $k_T$ and $l_T$, which are determined by the size of the nucleon
 ($R_N$) $ k_T \approx l_T \sim 1/R_N \gg\,Q_T ( |\vec{Q}_T \,+
\,\vec{p}_{T,12}|)  \sim 1/R_D$,  through the impact factors $I_P$
 in \eq{BAD1} and \eq{BAD2}.  The reason is that $R_D\,\,\gg\,\,R_N$.
  Neglecting $Q_T$ and $p_{T,12}$ in comparison with $k_T$ and $l_T$,
       we can simplify \eq{BAD1} and \eq{BAD2}  to the form  
      \bea\label{BAD4}
 & &\frac{d^2 \sigma}{d y_1 \,d y_2  d^2 p_{T1} d^2 p_{T2}}\Lb
 \fig{ddba}-a\Rb\,+\,\frac{d^2 \sigma}{d y_1 \,d y_2  d^2 p_{T1}
 d^2 p_{T2}}\Lb \fig{ddba}-b\Rb\,\propto\,\\
& & \frac{1}{p^2_{T1} \,p^2_{T2}}\,\int d^2 Q_T \Bigg\{ G^2_D\Lb
 Q_T\Rb \,+\,\frac{1}{2\Lb N^2_c - 1\Rb}\,G_D\Lb Q_T\Rb \,G_D\Lb
 \vec{Q}_T\,+\,\vec{p}_{T,12}\Rb\Bigg\}\nn\\
  &\times& \int d^2k_t\,d^2 l_T
\Bigg( \frac{I_P\Lb \vec{k}_T, -\vec{k}_T \Rb\,I_P\Lb \vec{k}_T -
 \vec{p}_{T1}, -\vec{k}_T   +\vec{p}_{T1}\Rb}{\vec{k}^2_T\,\Lb
 \vec{k}-\vec{p}_1\Rb^2_T}\Bigg)\,\times\,
\Bigg( \frac{I_P\Lb \vec{l}_T, -\vec{l}_T \Rb\,I_P\Lb \vec{l}_T
 - \vec{p}_{T2}, -\vec{l}_T   +\vec{p}_{T2}\Rb}{\vec{l}^2_T\,\Lb
 \vec{l}-\vec{p}_2\Rb^2_T} \Bigg)\nn\eea
      
         In \eq{BAD4} we consider $p_1 = p_2$ for the expressions in
 $\Big(\dots\Big)$.  Note, that the interference diagram
 of \fig{ddba}-b contributes when the polarizations of the produced gluons
 are the same,  this fact is reflected in \eq{BAD2} by the same indices
 of Lipatov vertices. In \eq{BAD4} we replace
      \bea \label{BAD5}       
&&\Gamma_\mu\Lb \vec{k}_T, \vec{p}_{T1}\Rb\,\Gamma_\mu\Lb -\vec{l}_T
 -\vec{Q}_T, \vec{p}_{T1}\Rb\,\Gamma_\mu\Lb \vec{l}_T, \vec{p}_{T2}\Rb\,
\Gamma_\mu\Lb -\vec{k}_T +\vec{Q}_T, \vec{p}_{T2}\Rb \nn\\
&&~~~~~~~~~~~~~~~~~~~~~~~~~~~~~~~~ \to~~~\h 
\Gamma_\mu\Lb \vec{k}_T, \vec{p}_{T1}\Rb   \Gamma_\mu\Lb -\vec{k}_T +\vec{Q}_T, \vec{p}_{T2}\Rb
\,   \Gamma_\nu\Lb \vec{l}_T, \vec{p}_{T2}\Rb   \Gamma_\nu\Lb -\vec{l}_T -\vec{Q}_T, \vec{p}_{T1}\Rb  \\
&&~~~~~~~~~~~~~~~~\xrightarrow{Q_T,p_{T,12} \,\ll\, k_t,l_T}~~~\h 
\Gamma_\mu\Lb \vec{k}_T, \vec{p}_{T1}\Rb   \Gamma_\mu\Lb -\vec{k}_T , \vec{p}_{T1}\Rb
\,   \Gamma_\nu\Lb \vec{l}_T, \vec{p}_{T2}\Rb   \Gamma_\nu\Lb -\vec{l}_T , \vec{p}_{T2}\Rb  \nn\\
&& ~~~~~~~~~~~~~~~~~~~~~~~~~~~~~~~~=~~~\,\,\,\,\h \frac{1}{p^2_{T1}\,p^2_{T2}}\,k^2_T\,\Lb \vec{k} - \vec{p}_1\Rb^2_T\,\,l^2_T\,\Lb \vec{l} - \vec{p}_2\Rb^2_T \label{BAD51}
\eea
  Factor $1/\Lb N^2_c - 1\Rb$ in \eq{BAD4} reflects that identical gluons
 have the same colors ($N_c$ is the number of colors).
  
  Finally, the correlation function $ C\Lb R_c |\vec{p}_{T1}\,-\,\vec{p}_{T2}|\Rb$ in \eq{I2} is equal to
  \beq \label{BAD6}
   C\Lb R_D\,p_{T,12}\Rb\,\,=\,\,\frac{1}{2 \Lb N^2_C - 1\Rb}\frac{\int d^2 Q_T\,G_D\Lb Q_T\Rb\,  G_D\Lb | \vec{Q}_T   
    \,+\,\vec{p}_{T,12}|\Rb}{\int d^2 Q_T\,G^2_D\Lb Q_T\Rb}\,    
    \eeq
    %%%%%%%%%%%%%%%%%%%%%%%%%%%%%%%%%%%%%%%%%%%%%%%
{\boldmath{\subsection{Bose-Einstein correlation function with radius
 $\propto\,\, R_N$: Glauber corrections}}}
%%%%%%%%%%%%%%%%%%%%%%%%%%%%%%%%%%%%%%%%%%%%%%% 

\begin{figure}[h]
 \includegraphics[width=16cm]{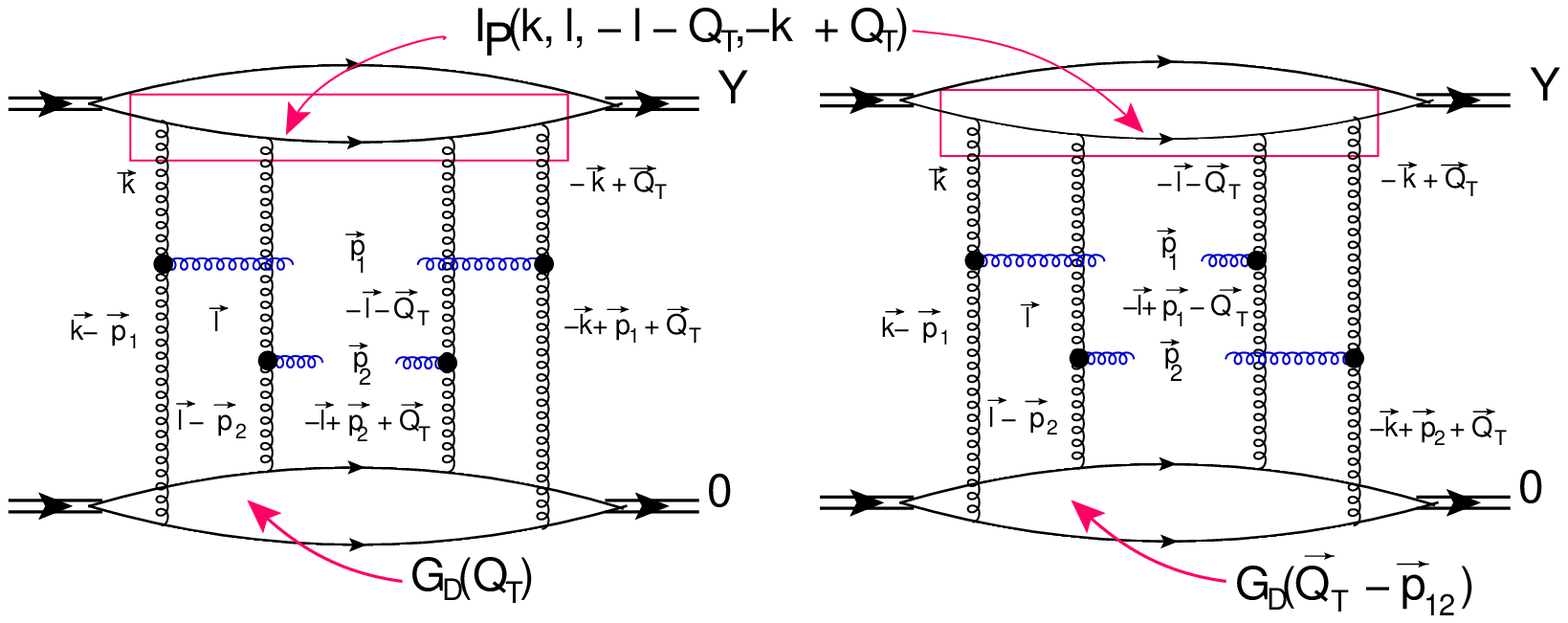}
   \protect\caption{ The double inclusive production of two gluons with
 rapidities $y_1$ and $y_2$ and transverse momenta $\vec{p}_{T1}$ and
 $\vec{p}_{T2}$ in the Born Approximation of perturbative QCD. The
 interference diagram of \fig{pdba}-b yields the correlation function
 $C\Lb R_N |\vec{p}_{T1}\,-\,\vec{p}_{T2}|\Rb$ of \eq{I2}. The 
 solid lines denote nucleons in the deuterons, which are  illustated 
by 
double lines}
\label{pdba}
   \end{figure}

In this subsection we  show that the Glauber corrections due to
 interaction of one nucleon with two nucleons of the deuteron, shown
 in \fig{pdba}, lead to a correlation radius of the order of $R_N$.
  In the diagram of \fig{pdba}-a, $Q_T$ is of the order of $1/R_D$ and,
 therefore, it is much smaller that the typical values of $k_T$ and $l_T$,
 which are of the order of $1/R_N$. Hence, the contribution of this diagram
 is similar to \eq{BAD4}: viz.
\bea \label{BAPD1}
 & &\frac{d^2 \sigma}{d y_1 \,d y_2  d^2 p_{T1} d^2 p_{T2}}\Lb \fig{pdba}-a\Rb\,\,\propto\,
 \frac{1}{p^2_{T1} \,p^2_{T2}}\,\int d^2 Q_T\,  G_D\Lb Q_T\Rb\nn\\
  &\times& \int d^2 k_T\,d^2 l_T\,I_P\Lb \vec{k}_T,\vec{l}_T, -\vec{l}_T , - \vec{k}_T\Rb
 \frac{I_P\Lb \vec{k}_T - \vec{p}_{T1}, -\vec{k}_T   +\vec{p}_{T1}\Rb\,I_P\Lb \vec{l}_T - \vec{p}_{T2}, -\vec{l}_T   +\vec{p}_{T2}\Rb}{\vec{k}^2_T\,\Lb \vec{k}-\vec{p}_1\Rb^2_T\,\vec{l}^2_T\,\Lb \vec{l}-\vec{p}_2\Rb^2_T}\nn\eea

 %%%%%%%%%%%%%%%%%%%%%%%%%%%%%%%%%%%%%%%%%%%%%%%%%%%%%%

\begin{figure}[ht]
 \includegraphics[width=14cm]{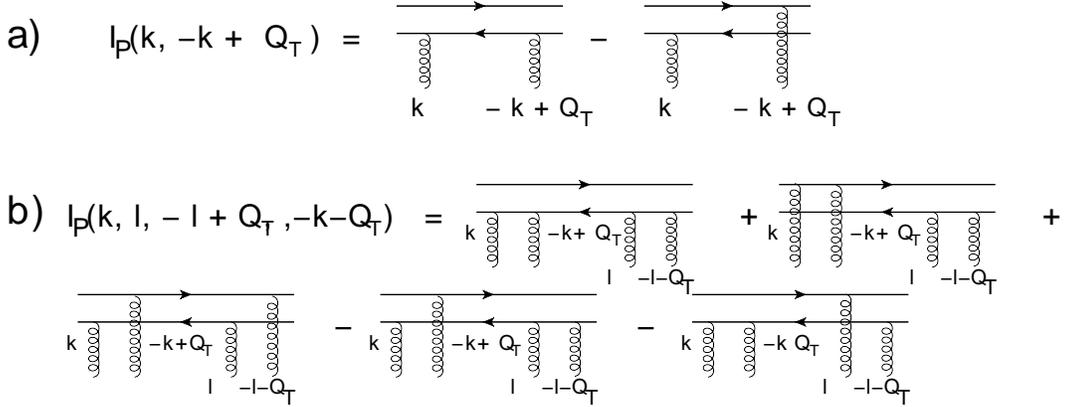}
   \protect\caption{ The impact factors for onium for $R_N k_T> 1$,
 $ R_N l_T> 1$ and $R_N Q_T> 1$ }
\label{if}
   \end{figure}

 %%%%%%%%%%%%%%%%%%%%%%%%%%%%%%%%%%%%%%%%%%%%%%%%%%%%%%
 Unfortunately, we cannot treat the impact factors $I_P$   theoretically
 in the case of  nucleon. The phenomenological approach to $I_P$ has been
 discussed in Refs.\cite{LERE,GLMBE}, and we will
 return to this below.  For the moment we replace the nucleon by the 
state
 of  a heavy quark and antiquark (onium), to study the key features of 
the
 impact factors in the framework of perturbative QCD (see \fig{if}).
 Introducing, the form factor of the onium in the form
 \beq \label{ONFF}
 F\Lb Q_T\Rb\,\,=\,\,\int d^2 r \,e^{i \h  \vec{Q}_T\cdot \vec{r}}\,| \Psi_{\rm onium}\Lb r\Rb|^2
 \eeq
 we can express the impact factor in the form
 \bea 
 I_P\Lb \vec{k}_T,-\vec{k}_T + \vec{Q}_T\Rb &=& F\Lb \vec{Q}_T\Rb\,-\,  F\Lb 2 \vec{k}_T + \vec{Q}_T \Rb;\label{IF1}\\
  I_P\Lb \vec{k}_T,\vec{l}_T, - \vec{l}_T + \vec{Q}_T,-\vec{k}_T+  \vec{Q}_T\Rb &=& 
  1\,+\, F\Lb 2 \vec{Q}_T\Rb\,+\, F\Lb 2 (\vec{k}_T + \vec{l}_T)\Rb\,+\,  F\Lb 2 (\vec{k}_T - \vec{l}_T - \vec{Q}_T)\Rb \nn\\
  &-&  F\Lb 2 \vec{k}_T \Rb \,-\, F\Lb 2 (\vec{k}_T - \vec{Q}_T)\Rb \,-\,F\Lb 2 \vec{l}_T \Rb \,-\,  F\Lb 2 (\vec{l}_T + \vec{Q}_T)\Rb  \label{IF2} \eea
  In \eq{BAPD1} the impact factors are equal to
  \bea
 \hspace{-1cm}  I_P\Lb \vec{k}_T,-\vec{k}_T\Rb &=& 1\,-\,  F\Lb 2 \vec{k}_T \Rb; ~~~~~~
   I_P\Lb \vec{l}_T,-\vec{l}_T\Rb \,=\, 1\,-\,  F\Lb 2 \vec{l}_T \Rb;\label{IF3}\\
\hspace{-1cm}  I_P\Lb \vec{k}_T,\vec{l}_T, - \vec{l}_T,-\vec{k}_T \Rb &=& 
 2\,+\, F\Lb 2 (\vec{k}_T + \vec{l}_T)\Rb +  F\Lb 2 (\vec{k}_T - \vec{l}_T )\Rb -  F\Lb 2 \vec{k}_T \Rb \,-\, F\Lb 2 (\vec{k}_T )\Rb - F\Lb 2 \vec{l}_T \Rb -  F\Lb 2 (\vec{l}_T )\Rb;  \label{IF4}
   \eea
  
   The integration over $k_T$ and $l_T$  lead to
 typical values of $k_T \sim 1/R_N$ and $l_T \sim 1/R_N$, and it 
 does not generate  azimuthal angle correlations.  Indeed,  this
 is clear from the following features of $I_P$ from \eq{IF4}:
  \bea \label{BAPD2}  R_N\,k_T \ll 1,  R_N\,l_T \ll 1;~~~~   I_P\Lb \vec{k}_T,\vec{l}_T, - \vec{l}_T,-\vec{k}_T \Rb\,&\propto&  \,k^2_T\,l^2_T;\nn\\
    R_N\,k_T \ll 1,  R_N\,l_T \gg  1;~~~~   I_P\Lb \vec{k}_T,\vec{l}_T, - \vec{l}_T,-\vec{k}_T \Rb\,&\propto&  \,k^2_T; \nn\\
     R_N\,k_T \gg 1,  R_N\,l_T \ll 1;~~~~   I_P\Lb \vec{k}_T,\vec{l}_T, - \vec{l}_T,-\vec{k}_T \Rb\,&\propto&  l^2_T; \nn\\
 R_N\,k_T \gg 1,  R_N\,l_T \gg 1;~~~~   I_P\Lb \vec{k}_T,\vec{l}_T, - \vec{l}_T,-\vec{k}_T \Rb\,&\propto&  1;
  \eea

   %%% %%%%%%%%%%%%%%%%%%%%%%%%%%%%%%%%%%%%%%%%%%%%%%%%%

 %%%%%%%%%%%%%%%%%%%%%%%%%%%%%%%%%%%%%%%%%%%%%%%%
  In the diagram of \fig{pdba}-b one can see that
 $G_D\Lb \vec{Q}_T - \vec{p}_{12}\Rb$  regulates that
  $| \vec{Q}_T - \vec{p}_{12}  |$
   is of the order of $1/R_D$. This means that we can put
 $\vec{Q}_T = \vec{p}_{12}$ in all parts of diagrams, since
   the typical values of $k_T$ and $l_T$  are about 
 $1/R_N \,\gg\,1/R_D$. Therefore, the diagram of \fig{pdba}-b
 can be reduced to the form
   \bea \label{BAPD3}
   & &\frac{d^2 \sigma}{d y_1 \,d y_2  d^2 p_{T1} d^2 p_{T2}}\Lb \fig{pdba}-b\Rb\,\,\propto\,
 \frac{1}{p^2_{T1} \,p^2_{T2}}\,\int d^2 Q_T G_D\Lb \vec{Q}_T - \vec{p}_{12}\Rb\nn\\
  &\times& \int d^2k_T\,d^2 l_T\,I_P\Lb \vec{k}_T,\vec{l}_T, -\vec{l}_T - \vec{p}_{12} , - \vec{k}_T +\vec{p}_{12}\Rb
  \,I_P\Lb \vec{k}_T - \vec{p}_{T1}, -\vec{k}_T   +\vec{p}_{T1}\Rb  \,I_P\Lb \vec{l}_T - \vec{p}_{T2}, -\vec{l}_T   + \vec{p}_{T2}\Rb\nn\\
  &\times& \Bigg\{\Bigg(\frac{1}{\Lb \vec{k} - \vec{p}_1\Rb^2_T\,\Lb \vec{l} + \vec{p}_{12}\Rb^2_T}\,+\,\frac{1}{k^2_T\,\Lb \vec{l} - \vec{p}_2\Rb^2_T}\Bigg)\,-\,\frac{\Lb \vec{l} - \vec{k} +\vec{p}_{12}\Rb^2_T\,p^2_{T,1}}{k^2_T\,\Lb \vec{k} - \vec{p}_1\Rb^2_T
   \,\Lb \vec{l} - \vec{p}_2\Rb^2_T \,\Lb \vec{l} + \vec{p}_{12}\Rb^2}\Bigg\}
  \nn\\
   &\times& \Bigg\{\Bigg(\frac{1}{\Lb \vec{l} - \vec{p}_2\Rb^2_T\,\Lb \vec{k} - \vec{p}_{12}\Rb^2_T}\,+\,\frac{1}{l^2_T\,\Lb \vec{k} - \vec{p}_1\Rb^2_T}\Bigg)\,-\,\frac{\Lb \vec{l} - \vec{k} +\vec{p}_{12}\Rb^2_T\,p^2_{T,2}}{l^2_T\,\Lb \vec{l} - \vec{p}_2\Rb^2_T
   \,\Lb \vec{k} - \vec{p}_1\Rb^2_T \,\Lb \vec{k} - \vec{p}_{12}\Rb^2}\Bigg\}  
  \eea   
  The largest contributions to the integrals over $k_T$ and $l_T$ lead to the
 logarithmically large terms, which are proportional to $\ln\Lb p^2_{T1}
\,R^2_N\Rb\,
  \ln\Lb p^2_{T2}\,R^2_N\Rb$. 
 These 
  contribution stems from the terms which are  proportional
 to $1/\Lb\Lb \vec{k} - \vec{p}_1\Rb^2_T\Rb^2$, and to  $1/\Lb\Lb
 \vec{l} - \vec{p}_2\Rb^2_T\Rb^2$. We consider the kinematic
 region in the integration over $k_T$ and $l_T$, where $\vec{k}_T
 -  \vec{p}_{T1} \equiv \vec{k}_1 \to 0 $ and $\vec{l}_T -  \vec{p}_{T2}
 \equiv \vec{l}_2 \to 0$. For small $k_1\,\ll\, p_{T2}$  and $l_2 \,\ll\, p_{T1} $ the product of curly brackets  is equal to
  \beq  \label{BAPD4}
  \frac{1}{p^2_{T1}} \Bigg\{\frac{1}{k^2_1} \,+\,\frac{1}{l^2_1} \,-\,\frac{\Lb \vec{k}_1\,-\,\vec{l}_2\Rb^2}{k^2_1\,l^2_2}\Bigg\}\,  \frac{1}{p^2_{T2}} \Bigg\{\frac{1}{k^2_1} \,+\,\frac{1}{l^2_1} \,-\,\frac{\Lb \vec{k}_1\,-\,\vec{l}_2\Rb^2}{k^2_1\,l^2_2}\Bigg\}  =\,\frac{4 \Lb \vec{k}_1 \cdot \vec{l}_2\Rb^2}{p^2_{T1}\,p^2_{T2}\,k^4_1\,l^4_2}\,\xrightarrow{\rm after \,integration\,over\,angle}\,\frac{2}{p^2_{T1}\,p^2_{T2}\,k^2_1\,l^2_2}
  \eeq
  
  For $ R_N k_1 \ll 1$ $,I_P\Lb \vec{k}_T - \vec{p}_{T1}, -\vec{k}_T
   +\vec{p}_{T1}\Rb  \propto k^2_1$, and the integral over $k_1$ gives
  a small contribution of the order of $1/(R^2_N p^2_{T1})$. Recall 
that
 we can use the perturbative QCD approach only if $R_N p_{T1} \,\gg\,1
 $ and $R_N p_{T2} \,\gg\,1 $. Therefore, the main contribution occurs
 from  the region of integration $(1/R^2_N) \ll k_1 \ll p_{T2}$ and
 $(1/R^2_N ) \ll l_1 \ll p_{T1}$. Integration in this region leads to
 the contribution
  \bea \label{BAPD5}
\frac{d^2 \sigma}{d y_1 \,d y_2  d^2 p_{T1} d^2 p_{T2}}\Lb \fig{pdba}-b\Rb\,\,&\propto\,&
 \frac{1}{p^4_{T1} \,p^4_{T2}}\,\ln\Lb p^2_{T1} R^2_N\Rb\ln\Lb p^2_{T2} R^2_N\Rb \int d^2 Q_T G_D\Lb \vec{Q}_T - \vec{p}_{12}\Rb\nn\\  
&\times & I_P\Lb \vec{p}_{T1},\vec{p}_{T2},  -\vec{p}_{T1} ,- \vec{p}_{T2}\Rb
 \eea
 Therefore, from this kinematic region the correlations are determined by
 the impact factor.
   Using the impact factor given in \eq{IF2}, we see that 
  \beq \label{BAPD6}
   I_P\Lb \vec{p}_{T1},\vec{p}_{T2},  \vec{p}_{T1} , \vec{p}_{T2}\Rb\,=\,
   2 + F\Lb 2\Lb \vec{p}_{T1} + \vec{p}_{T2}\Rb\Rb + F\Lb 2 \vec{p}_{T,12}\Rb - 2 F\Lb 2 \vec{p}_{T1}\Rb - 2 F\Lb \vec{p}_{T2}\Rb
   \eeq
   This function is symmetric with respect to $\phi \to 
 \pi -\phi$ ($\vec{p}_{T1} \to - \vec{p}_{T1}$),  and with such an
 impact factor, the Born approximation produces only $v_n$ with even $n$, 
as 
was
 noted in Ref.\cite{KOWE,KOLUCOR}. However, this conclusion is based on the
 impact factor of \eq{IF2}. \eq{BAPD6} shows that this impact factor leads
 to $p_{T,12} \sim 1/R_N$.

       %%% %%%%%%%%%%%%%%%%%%%%%%%%%%%%%%%%%%%%%%%%%%%%%%%%%
\begin{figure}[ht]
 \includegraphics[width=15cm]{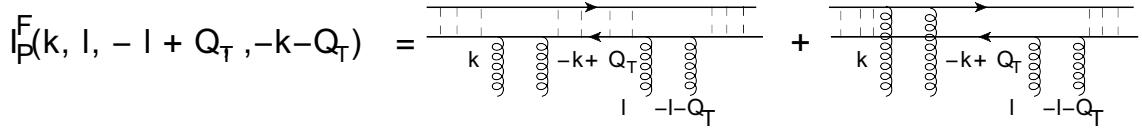}
   \protect\caption{ The impact factors for onium for  for $R_N k_T> 1$,
 $R_N l_T> 1$ and $R_N Q_T\leq  1$. The  dashed lines denote the 
Coulomb
 gluons that form the onium bound state. }
\label{iff}
   \end{figure}

 %%%%%%%%%%%%%%%%%%%%%%%%%%%%%%%%%%%%%%%%%%%%%%%%    
    
     Note that the simple expression of \eq{IF2} (see \fig{if})
 is written for sufficiently hard gluons. For small values of $Q_T =
 p_{T,12}$ we need to add the first diagram of \fig{iff}, in which two
 gluons with large transverse momenta (about $p_{T1}$ or $p_{T2}$ ) but
 small $Q_T$. The final expression for the impact factor takes the form
    
    \beq \label{IFF}
      I^F_P\Lb \vec{k}_T,\vec{l}_T, - \vec{l}_T + \vec{Q}_T,-\vec{k}_T+  \vec{Q}_T\Rb    \,=\,  I_P\Lb \vec{k}_T, - \vec{k}_T + \vec{Q}_T   \Rb  \,I_P\Lb \vec{l}_T, - \vec{l}_T - \vec{Q}_T   \Rb   \,+\, \underbrace{  I_P\Lb \vec{k}_T,\vec{l}_T, - \vec{l}_T + \vec{Q}_T,-\vec{k}_T+  \vec{Q}_T\Rb }_{\rm \eq{IF2}}
      \eeq 
  The first term in \eq{IFF} generates the correlation function which is
 proportional to $F^2\Lb 2 \vec{p}_{T,12}\Rb $.  
  
   Summarizing, we see that the Born approximation of perturbative 
QCD, generates
 the correlation function which is determined by the impact factor of the
 nucleon, the typical correlation length is about  $R_N$, and even for 
the
 unrealistic perturbative model of onium, this correlation function is not
 symmetric with respect to $\phi \to \pi - \phi$. We will consider below the  
more realistic case, 
  in leading log approximation of perturbative QCD.
  However, we would like to stress  now, that  the correlation function
 stems from the large non-perturbative distances of the order of the 
nucleon size.

         %%%%%%%%%%%%%%%%%%%%%%%%%%%%%%%%%%%%%%%%%%%%%%%
{\boldmath{\subsection{Bose-Einstein correlation function
 with radius $\propto\,\, R_N$: nucleon-nucleon interaction}}}
%%%%%%%%%%%%%%%%%%%%%%%%%%%%%%%%%%%%%%%%%%%%%%% 

\begin{figure}[ht]
 \includegraphics[width=16cm]{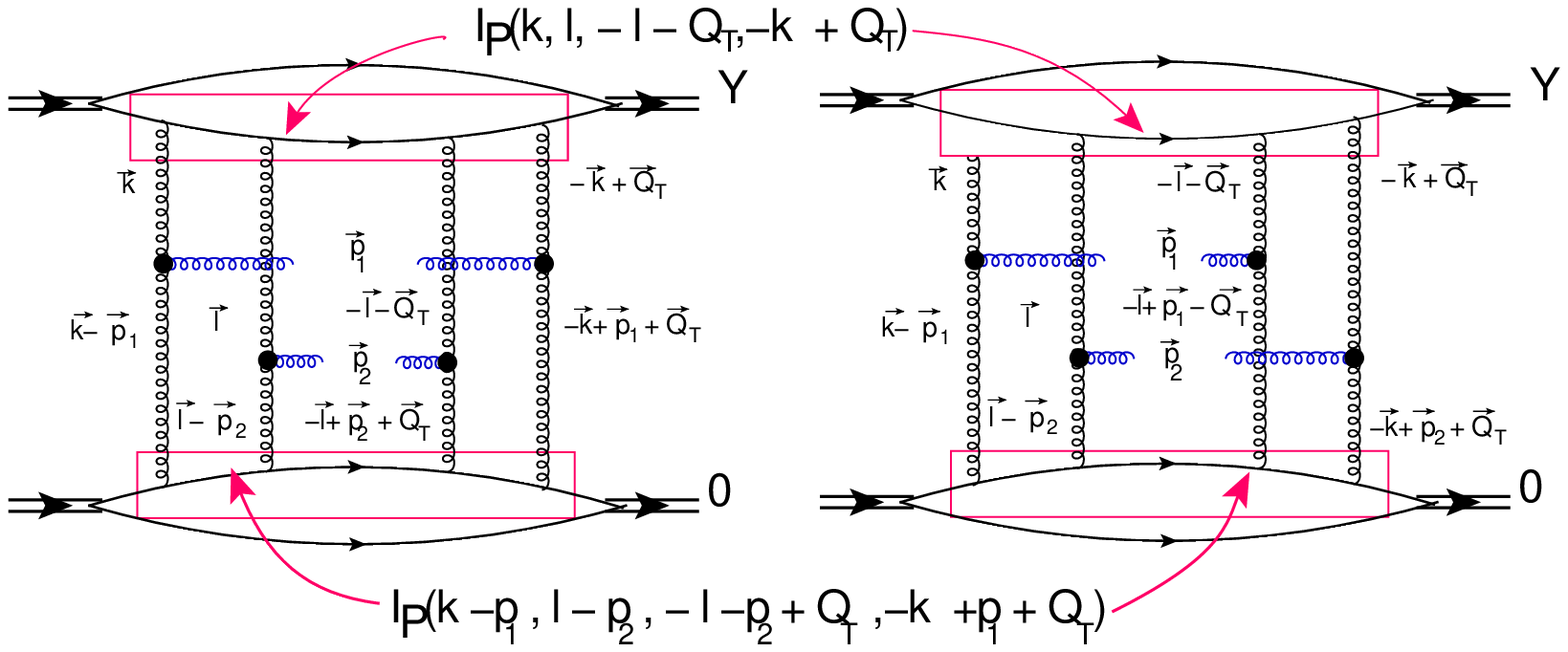}
   \protect\caption{ The double inclusive production of two gluons
 with rapidities $y_1$ and $y_2$ and transverse momenta $\vec{p}_{T1}$
 and $\vec{p}_{T2}$ in the Born Approximation of perturbative QCD for the
 nucleon-nucleon interaction. The interference diagram of \fig{ppba}-b
 yields the correlation function $C\Lb R_D 
|\vec{p}_{T1}\,-\,\vec{p}_{T2}|\Rb$
 of \eq{I2}. The  solid lines denote nucleons in the deuterons, which
 are  illustrated by double lines.}
\label{ppba}
   \end{figure}

 The correlations with $R_c = R_N$ are typical for the 
nucleon-nucleon
 interaction ( see \fig{ppba} for the Born approximation of perturbative
 QCD). However, we will consider them below for the general case of
 the production of two parton showers, since we prefer to use a more
 phenomenological and realistic approach for the impact factors
 $I_P$, than we explored  above, replacing the nucleon by the onium 
state.

   %%%%%%%%%%%%%%%%%%%%%%%%%%%%%%%%%%%%%%%%%%%%%%%%%%%%%%

 %%%%%%%%%%%%%%%%%%%%%%%%%%%%%%%%%%%%%%%%%%%%%%%%
  %%%%%%%%%%%%%%%%%%%%%%%%%%%%%%%%%%%%%%%%%%%%%%%%
   \section{Production of two parton showers} 
   %%%%%%%%%%%%%%%%%%%%%%%%%%%%%%%%%%%%%%%%%%%%%%%
{\boldmath{\subsection{ $ R_c \,\,\propto\,\, R_D$}}}
%%%%%%%%%%%%%%%%%%%%%%%%%%%%%%%%%%%%%%%%%%%%%%% 
In this section we consider the general case of the production of two parton
 showers  shown in \fig{ddagk}.  In  the leading log approximation
 (LLA) of perturbative QCD, the structure of one parton shower is 
described
 by the BFKL Pomeron\cite{BFKL,LI}.
  %%%%%%%%%%%%%%%%%%%%%%%%%%%%%%%%%%%%%%%%%%%%%%%%%%%%%%
\begin{figure}[ht]
 \includegraphics[width=16cm]{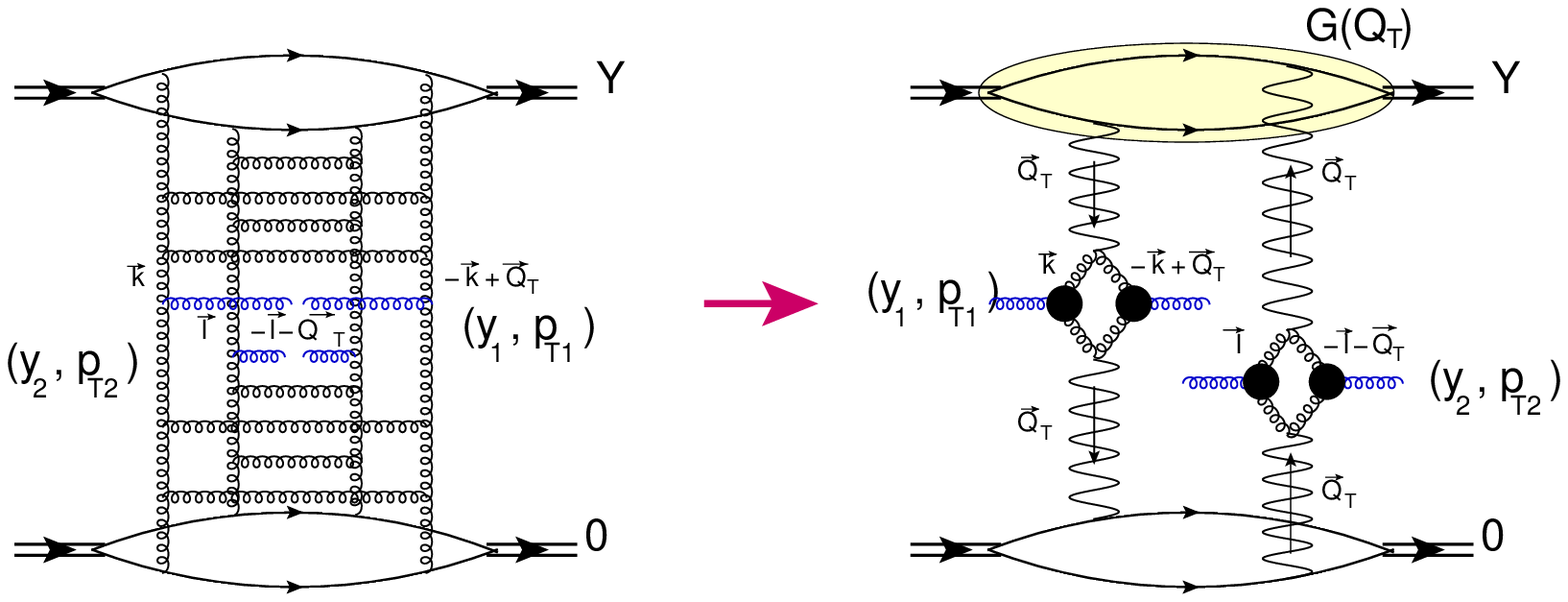}
   \protect\caption{ The double inclusive production of two gluons
 with rapidities $y_1$ and $y_2$ and transverse momenta $\vec{p}_{T1}$
 and $\vec{p}_{T2}$,  for the exchange of two BFKL Pomerons which are
 denoted by  wavy lines. This diagram is the LLA generalization of
 \fig{ddba}-a.
 The  solid lines denote nucleons in the deuterons, which are
 indicated by double lines.}
\label{ddspom}
   \end{figure}
 %%%%%%%%%%%%%%%%%%%%%%%%%%%%%%%%%%%%%%%%%%%%%%%%    
 In the leading log approximation of perturbative
 QCD the Born diagram of \fig{ddba}-a can be generalized to \fig{ddspom}.
 The contribution of  this diagram can be written as follows
\bea \label{RD1}
& &\frac{d^2 \sigma}{d y_1 \,d y_2  d^2 p_{T1} d^2 p_{T2}}\Lb \fig{ddspom}\Rb\,\,=\,\,\Lb\frac{2 \pi \bas}{C_F}\Rb^2  \frac{1}{p^2_{T,1}\,p^2_{T2}}\int d^2 Q_T  G^2_D\Lb Q_T\Rb \nn\\
&\times&\,\Bigg(\int d^2 k_{T}\,\,\phi^{N}_G\Lb Y - y_1; \vec{k}_T, -\vec{k}_T + \vec{Q}_T\Rb\,\phi^{N}_G\Lb y_1; \vec{k}_T- \vec{p}_{T1},-\vec{k}_T+ \vec{p}_{T1} +\vec{Q}_T \Rb\Bigg)\nn\\
&\times&\,\Bigg(\int d^2 l_{T}\,\,\phi^{N}_G\Lb Y - y_2; \vec{l}_T, -\vec{l}_T - \vec{Q}_T\Rb\,\phi^{N}_G\Lb y_2; \vec{l}_T- \vec{p}_{T2},-\vec{l}_T+ \vec{p}_{T2} -\vec{Q}_T \Rb\Bigg)\eea

where $\phi^{N}_G\Lb y, \vec{k}_T,-\vec{k}_T + \vec{Q}_T\Rb$  denotes 
the
 probability to find a gluon with rapidity $ y $   and transverse 
momentum
$k_\perp$,
  in the process with momentum transferred  $Q_T$. In
 \eq{RD1} $ \bas \,= \,\as N_c/\pi$ with the number of colours equal to 
$N_c$.
  $\phi^{N}_G$ are the solutions of the BFKL  evolution equation 
\bea \label{RD2}
\frac{\partial \phi^N_G\Lb y, \vec{k}_T, -\vec{k}_T + \vec{Q}_T \Rb}{\partial y}\,&=&\,\bas\int \frac{d^2 k'_T}{2 \pi}\,K\Lb Q_T; k_T,k'_T\Rb\,\phi^N_G\Lb y, \vec{k'}_T,-\vec{k'}_T + \vec{Q}_T\Rb\,\,\\
&-&\,\,\Lb \omega_G\Lb \vec{Q}_T - \vec{k}_T\Rb\,+\,\omega_G\Lb \vec{k}_T\Rb\Rb \phi^N_G\Lb y, \vec{k}_T,-\vec{k}_T + \vec{Q}_T\Rb\nn
\eea

where

\bea \label{RD3}
K\Lb Q_T,k_T,k'_T\Rb\,&=&\, \frac{1}{\Lb \vec{k}_T - \vec{k'}_{T}\Rb^2}\Bigg\{ \frac{k^2_T}{k'^2_T} \,+\,\frac{\Lb \vec{Q}_T - \vec{k}_T\Rb^2}{\Lb \vec{Q}_T - \vec{k'}_T\Rb^2}\,\,-\,\,\frac{\Lb \vec{k}_T - \vec{k'}_{T}\Rb^2}{k'^2_T\, \Lb\vec{Q}_T - \vec{k'}_T\Rb^2}\Bigg\}\\
\omega_G\Lb\vec{k}_T\Rb&=& \h \bas k^2_T \int \frac{d^2 k'_T}{2 \pi} \frac{1}{k'^2_T\,\Lb \vec{k}_T - \vec{k'}_{T}\Rb^2 }\nn
\eea

The typical momenta in $\phi^N_G$ is about $1/R_N$  or larger,
 (about  $p_{T1}$($p_{T2}$) or $Q_s$, where $Q_s$ denotes   the 
saturation
 scale. Bearing this in mind,  and noting that $Q_T \sim 1/R_D 
\,\ll\,1/R_N$
 we can put $Q_T  = 0$, in the arguments of $\phi^N_G$.  This simplifies 
\eq{RD1}
 reducing it to the following expression
\beq \label{RD4}
\frac{d^2 \sigma}{d y_1 \,d y_2  d^2 p_{T1} d^2 p_{T2}}\Lb
 \fig{ddspom}\Rb\,\,=\,\,
\frac{d^2 \sigma}{d y_1 \ d^2 p_{T1}}\,\frac{d^2 \sigma}{d y_2
 \ d^2 p_{T2}} \,\times\,\int d^2 Q_T  G^2_D\Lb Q_T\Rb
\eeq

\begin{figure}[ht]
 \includegraphics[width=16cm]{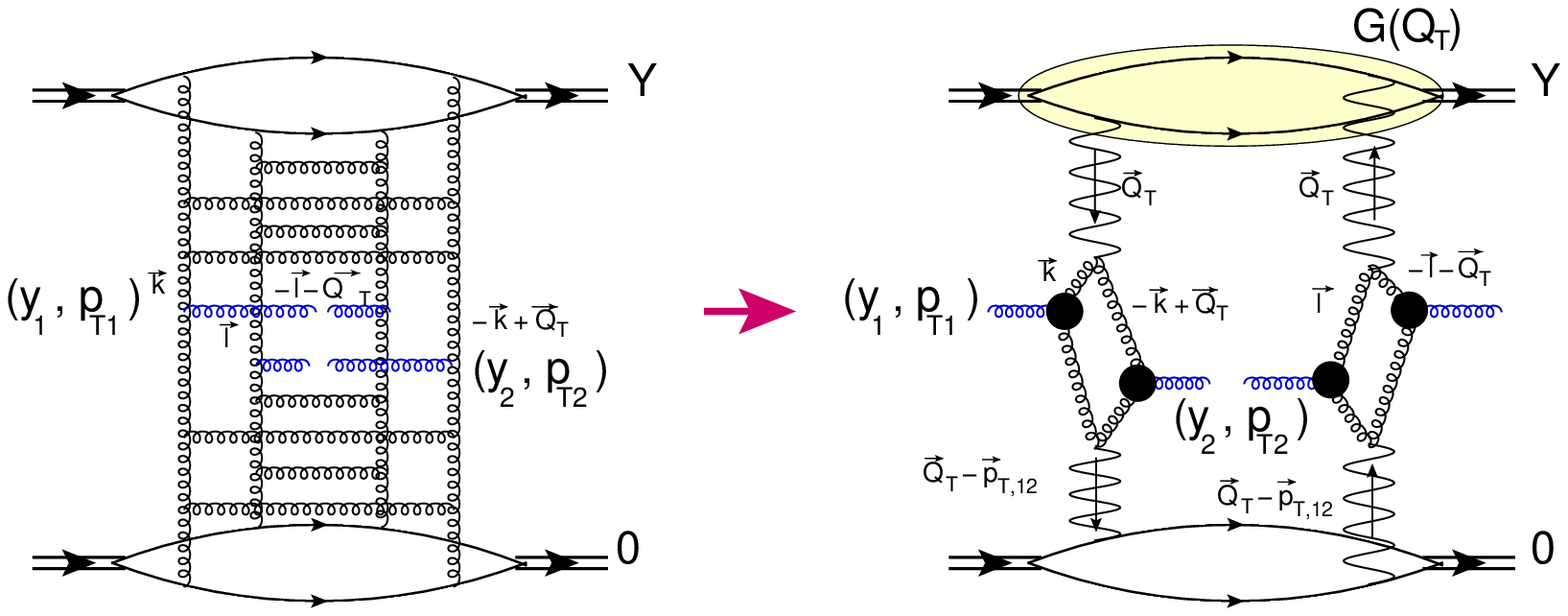}
   \protect\caption{ The double inclusive production of two gluons
 with rapidities $y_1$ and $y_2$ and transverse momenta $\vec{p}_{T1}$
 and $\vec{p}_{T2}$  for the exchange of two BFKL Pomerons which are
 denoted by  wavy lines. This diagram is the LLA generalization of 
\fig{ddba}-b.
 The  solid lines denote nucleons in the deuterons, which are 
represented by double lines.}
\label{ddipom}
   \end{figure}

The diagram of \fig{ddba}-b in the LLA, simplifies the expression for the 
exchange of two BFKL Pomerons, 
but with more complicated vertices. Using \eq{BAD5} and
 considering $\bas\Lb y_1- y_2 \Rb \leq 1$, we can write
 this exchange in the form that is represented in \fig{ddipom},
 and its contribution has the following form

  %%%%%%%%%%%%%%%%%%%%%%%%%%%%%%%%%%%%%%%%%%%%%%%%%%%%%%
 %%%%%%%%%%%%%%%%%%%%%%%%%%%%%%%%%%%%%%%%%%%%%%%%
\bea \label{RD5}
& &\frac{d^2 \sigma}{d y_1 \,d y_2  d^2 p_{T1} d^2 p_{T2}}\Lb \fig{ddipom}\Rb\,\,=\,\,\h \Lb\frac{2 \pi \bas}{C_F}\Rb^2  \int d^2 Q_T  G_D\Lb Q_T\Rb\,G_D\Lb \vec{Q}_T -  \vec{p}_{T,12}\Rb \\
&\times&\,\Bigg(\int d^2 k_{T}\,\,\phi^{N}_G\Lb Y - y_1; \vec{k}_T, -\vec{k}_T + \vec{Q}_T\Rb\,
  \Gamma_\mu\Lb k_T, p_{T1}\Rb\,  \Gamma_\mu\Lb - \vec{k}_{T} + \vec{Q}_T, p_{T2}\Rb\,
\phi^{N}_G\Lb y_2; \vec{k}_T- \vec{p}_{T1},-\vec{k}_T+ \vec{p}_{T2} +\vec{Q}_T \Rb\Bigg)\nn\\
&\times&\,\Bigg(\int d^2 l_{T}\,\,\phi^{N}_G\Lb Y - y_1; \vec{l}_T, -\vec{l}_T - \vec{Q}_T\Rb\,  \Gamma_\mu\Lb l_T,p_{T1}\Rb\,  \Gamma_\mu\Lb-\vec{ l}_{T} -  \vec{Q}_T,p_{T2}\Rb\phi^{N}_G\Lb y_2; \vec{l}_T- \vec{p}_{T2},-\vec{l}_T+ \vec{p}_{T1} -\vec{Q}_T \Rb\Bigg)\nn\eea

\begin{figure}[ht]
 \includegraphics[width=16cm]{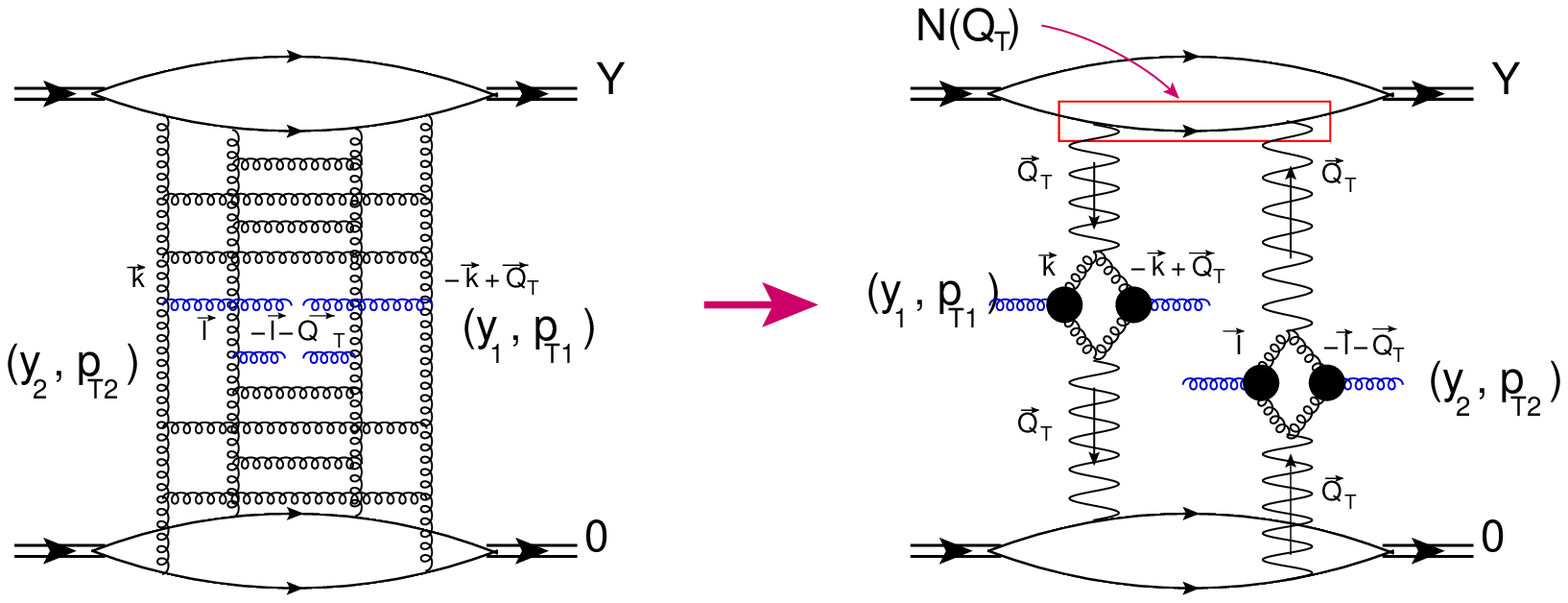}
   \protect\caption{ The  Mueller diagram\cite{MUDI} for the double
 inclusive production of two gluons with rapidities $y_1$ and $y_2$
 and transverse momenta $\vec{p}_{T1}$ and $\vec{p}_{T2}$,  for the
 exchange of two BFKL Pomerons which are denoted by wavy lines. This
 diagram is the LLA generalization of \fig{pdba}-a.
 The  solid lines denote nucleons in the deuterons, which are
 illustrated by double lines.}
\label{pdspom}
   \end{figure}

\begin{figure}[ht]
 \includegraphics[width=16cm]{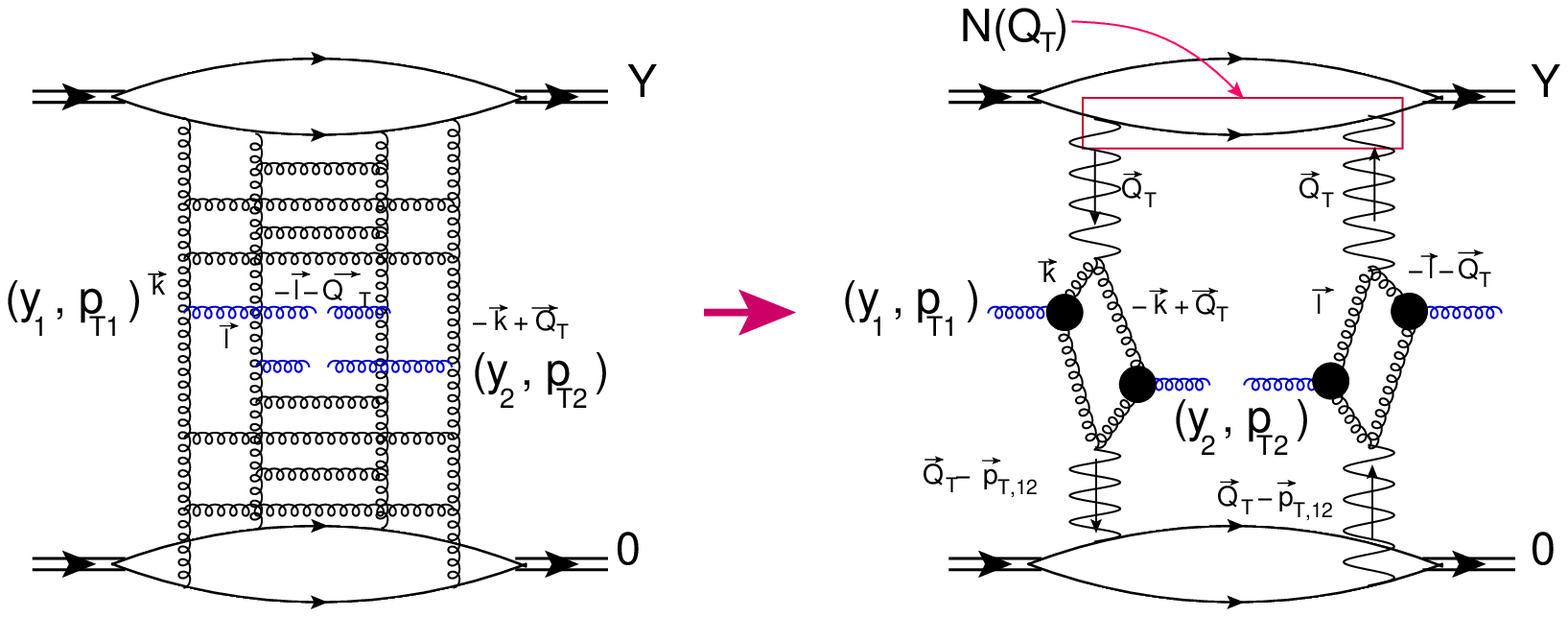}
   \protect\caption{ The Mueller diagram for the double inclusive production
 of two gluons with rapidities $y_1$ and $y_2$ and transverse momenta
 $\vec{p}_{T1}$ and $\vec{p}_{T2}$,  for the exchange of two BFKL Pomerons
 which are denoted by wavy lines. This diagram is the LLA generalization
 of \fig{pdba}-b.
 The  solid lines denote nucleons in the deuterons, which are 
illustated
 by double lines.}
\label{pdipom}
   \end{figure}

 Since $Q_T \sim 1/R_B \,\ll\,1/R_N$ as well as $|\vec{Q}_T - \vec{p}_{T,12}|
 \sim 1/R_D\,\ll\,1/R_N$, we can take both $Q_T= 0$ and $p_{T,12}=0$, but 
it is
 not  sufficient to reduce \eq{RD5} to \eq{BAD6}.  In addition we 
need to assume 
that
 $\bas\Lb y_1 - y_2\Rb\,\leq\, 1$.   Making this assumption we can 
replace
 $y_2$ in  $ \phi^{N}_G\Lb y_2; \vec{k}_T- \vec{p}_{T1},-\vec{k}_T+
 \vec{p}_{T2} +\vec{Q}_T \Rb $ by $y_1$ and $Y - y_1$, in $\phi^{N}_G\Lb
 Y - y_1; \vec{l}_T, -\vec{l}_T - \vec{Q}_T\Rb $ by $Y - y_2$. After these 
changes 
  \eq{RD5} can be reduced to the following expression
 \beq \label{RD6}
\frac{d^2 \sigma}{d y_1 \,d y_2  d^2 p_{T1} d^2 p_{T2}}\Lb \fig{ddipom}\Rb\,\,=\,\,
\frac{d^2 \sigma}{d y_1 \ d^2 p_{T1}}\,\frac{d^2 \sigma}{d y_2 \ d^2 p_{T2}} \,\times\,\int d^2 Q_T  G_D\Lb Q_T\Rb\,G_D\Lb \vec{Q}_T + \vec{p}_{T,12}\Rb
\eeq
 
 \eq{RD4} and \eq{RD6} lead to the same correlation
 function ($  C\Lb R_D\,p_{T,12}\Rb$)  \eq{BAD6} as in the Born approximation.
  %%%%%%%%%%%%%%%%%%%%%%%%%%%%%%%%%%%%%%%%%%%%%%%
{\boldmath{\subsection{ $ R_c \,\,\propto\,\, R_N$}}}
%%%%%%%%%%%%%%%%%%%%%%%%%%%%%%%%%%%%%%%%%%%%%
In LLA the diagrams of the Born approximation of \fig{pdba} can be
 generalized in the same way as has been discussed above. \fig{pdba}-a
 takes the form of \fig{pdspom} while the interference diagram of \fig{pdba}-b
 becomes \fig{pdipom}.

  %%%%%%%%%%%%%%%%%%%%%%%%%%%%%%%%%%%%%%%%%%%%%%%%%%%%%%
\begin{figure}[h]
 \includegraphics[width=11cm]{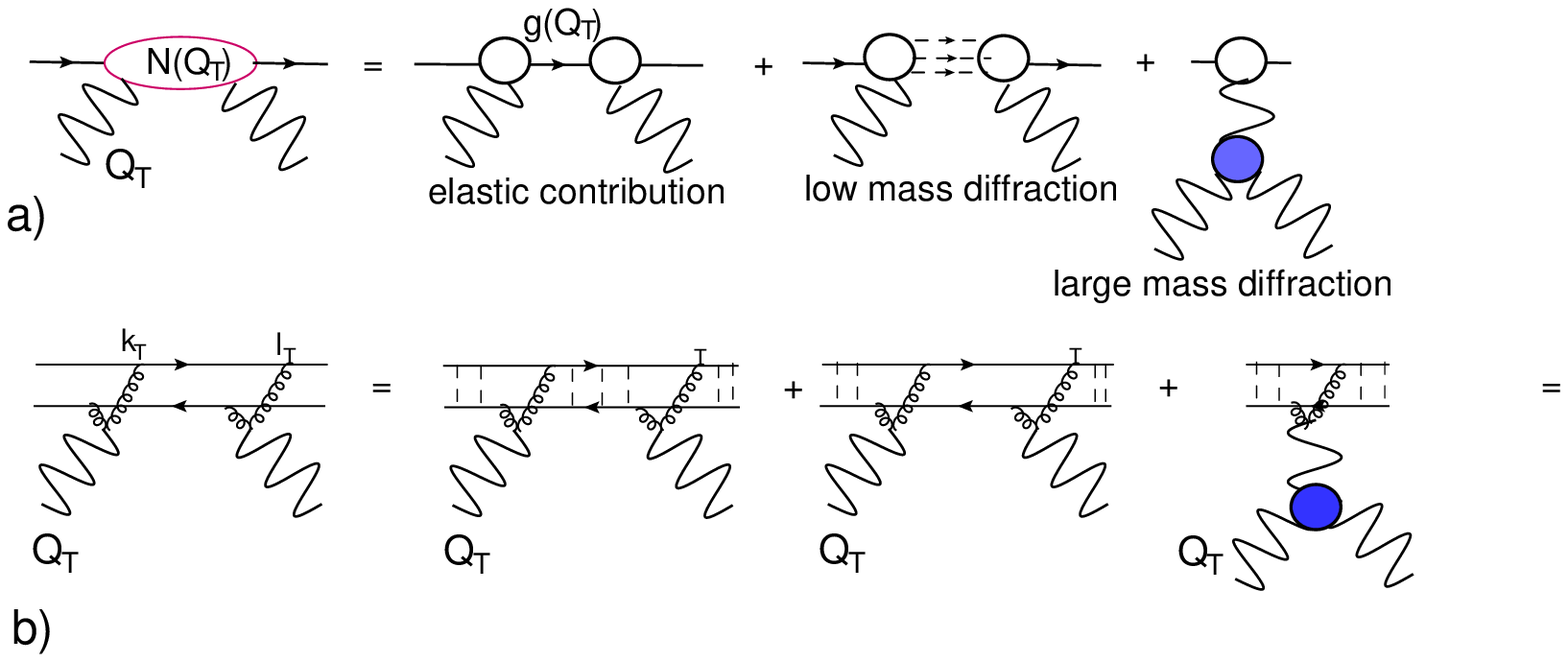}
   \protect\caption{ The structure of the amplitude $N\Lb Q_T\Rb$: \fig{n}-a
 for the BFKL Pomeron-nucleon interaction and \fig{n}-b for the BFKL
 Pomeron-onium interactions. The blob shows the triple BFKL Pomeron
 vertex which is the same for both figures. The  dashed vertical 
lines
 describe the Coulomb gluons that create the bound state: onium.}
\label{n}
   \end{figure}

 %%%%%%%%%%%%%%%%%%%%%%%%%%%%%%%%%%%%%%%%%%%%%%%%  
 The contribution of  the diagram of \fig{pdspom} can be written as follows
\bea \label{RN1}
& &\frac{d^2 \sigma}{d y_1 \,d y_2  d^2 p_{T1} d^2 p_{T2}}\Lb \fig{pdspom}\Rb\,\,=\,\,\Lb\frac{2 \pi \bas}{C_F}\Rb^2  \frac{1}{p^2_{T,1}\,p^2_{T2}}\int d^2 Q_T\,N\Lb Q_T\Rb\,  G^2_D\Lb Q_T\Rb \nn\\
&\times&\,\Bigg(\int d^2 k_{T}\,\,\phi^{N}_G\Lb Y - y_1; \vec{k}_T, -\vec{k}_T + \vec{Q}_T\Rb\,\phi^{N}_G\Lb y_1; \vec{k}_T- \vec{p}_{T1},-\vec{k}_T+ \vec{p}_{T1} +\vec{Q}_T \Rb\Bigg)\nn\\
&\times&\,\Bigg(\int d^2 l_{T}\,\,\phi^{N}_G\Lb Y - y_2; \vec{l}_T, -\vec{l}_T
 - \vec{Q}_T\Rb\,\phi^{N}_G\Lb y_2; \vec{l}_T- \vec{p}_{T2},-\vec{l}_T+ \vec{p}_{T2} -\vec{Q}_T \Rb\Bigg)\eea
where $N\Lb Q_T\Rb$ denotes the integral over all energies of the
 imaginary part of the Pomeron-nucleon scattering amplitude. This
 amplitude was introduced in Gribov's Pomeron calculus\cite{GRIBPOM},
 but it has been proven that we can use this formalism in LLA of
 perturbative QCD\cite{GLR}.  $N\Lb Q_T\Rb$  has the following
 general form(see \fig{n})
\beq \label{N}
N\Lb Q_T\Rb\,\,=\,\,\underbrace{g^2\Lb Q_T\Rb}_{\rm elastic\, scattering}\,
+\,\underbrace{\sum^{M_0}_{M_i=m}g^2\Lb Q_T; M_i\Rb}_{\rm diffraction\, in\,
 low\, masses}\,+\,\underbrace{\int_{M_0} \frac{d M^2}{M^2}\phi^N_G\Lb y_M,
 Q_T=0; \{\dots\}\Rb\,G_{3 \pom}\Lb Q_T,\{\dots\}\Rb}_{diffraction\, in\,
 high\, masses}\eeq
where $G_{3 \pom} $ is the triple BFKL Pomeron vertex, and $\{\dots\}$ 
denotes
 all  transverse momenta which we need to integrate over. $y_M =
 \ln\Lb M^2/M^2_0\Rb$.

\fig{n}-b shows how all contributions correspond to the onium case,
 where we can use perturbative QCD for theoretical estimates.

  %%%%%%%%%%%%%%%%%%%%%%%%%%%%%%%%%%%%%%%%%%%%%%%%%%%%%%
 %%%%%%%%%%%%%%%%%%%%%%%%%%%%%%%%%%%%%%%%%%%%%%%%
 
 Since $Q_T \sim 1/R_D \,\ll\,1.R_N$, and all other transverse momenta
 in \fig{pdspom}  are either of the order of $1/R_N$ or larger ( of 
the order
 of  $p_{T1}$,$p_{T2}$ or $Q_s$, where $Q_s$ is the saturation scale), we
 can safely put $Q_T = 0$ and reduce this contribution to the factorized form:
 \beq \label{RN2}
\frac{d^2 \sigma}{d y_1 \,d y_2  d^2 p_{T1} d^2 p_{T2}}\Lb \fig{pdspom}\Rb\,\,=
\frac{d^2 \sigma}{d y_1 \, d^2 p_{T1} }\,\frac{d^2 \sigma}{d y_2  \, d^2 p_{T2}}\,N\Lb Q_T=0\Rb\,\int d^2 Q_T G_D\Lb Q_T\Rb
\eeq

The contribution of the  relevant diagram, which is shown in 
\fig{pdipom}, can be written in the form:
\bea \label{RN3}
& &\frac{d^2 \sigma}{d y_1 \,d y_2  d^2 p_{T1} d^2 p_{T2}}\Lb \fig{pdipom}\Rb\,\,=\,\,\h \Lb\frac{2 \pi \bas}{C_F}\Rb^2  \int d^2 Q_T  N\Lb Q_T\Rb\,G_D\Lb \vec{Q}_T -  \vec{p}_{T,12}\Rb\nn \\
&\times&\,\Bigg(\int d^2 k_{T}\,\,\phi^{N}_G\Lb Y - y_1; \vec{k}_T, -\vec{k}_T + \vec{Q}_T\Rb\,
  \Gamma_\mu\Lb k_T, p_{T1}\Rb\,  \Gamma_\mu\Lb - \vec{k}_{T} + \vec{Q}_T, p_{T2}\Rb\,
\phi^{N}_G\Lb y_2; \vec{k}_T- \vec{p}_{T1},-\vec{k}_T+ \vec{p}_{T2} +\vec{Q}_T \Rb\Bigg)\nn\\
&\times&\,\Bigg(\int d^2 l_{T}\,\,\phi^{N}_G\Lb Y - y_1; \vec{l}_T, -\vec{l}_T - \vec{Q}_T\Rb\,  \Gamma_\mu\Lb l_T,p_{T1}\Rb\,  \Gamma_\mu\Lb-\vec{ l}_{T} -  \vec{Q}_T,p_{T2}\Rb\phi^{N}_G\Lb y_2; \vec{l}_T- \vec{p}_{T2},-\vec{l}_T+ \vec{p}_{T1} -\vec{Q}_T \Rb\Bigg)\eea

 Integration over $Q_T$ leads to $\vec{Q}_T - \vec{p}_{T,12}
 \sim 1/R_D\,\ll\,\,1/R_N$ and, therefore, as in the Born approximation
 we can put $\vec{Q}_T = \vec{p}_{T,12}$. In \eq{RN3} we have two sources
 of $p_{T,12}$ behavior: $N\Lb p_{T,12}\Rb $ and $\phi^N_G$. Replacing
 $\vec{Q}_T = \vec{p}_{T,12} $ we obtain
 
 \bea \label{RN4}
& &\frac{d^2 \sigma}{d y_1 \,d y_2  d^2 p_{T1} d^2 p_{T2}}\Lb \fig{pdipom}\Rb\,\,=\,\,\h \Lb\frac{2 \pi \bas}{C_F}\Rb^2  \int d^2 Q_T  N\Lb p_{T,12}\Rb\,G_D\Lb \vec{Q}_T -  \vec{p}_{T,12}\Rb\nn \\
&\times&\,\Bigg(\int d^2 k_{T}\,\,\phi^{N}_G\Lb Y - y_1; \vec{k}_T, -\vec{k}_T +\vec{p}_{T,12} \Rb\,  \frac{1}{k^2_T\,\Lb\vec{k}_T- \vec{p}_{T,12}\Rb^2 \Lb\vec{k}_T - \vec{p}_{T,1}\Rb^4}  \phi^{N}_G\Lb y_2; \vec{k}_T- \vec{p}_{T1},-\vec{k}_T+ \vec{p}_{T1} \Rb\nn\\
&\times&
  \Bigg\{\frac{\Lb\vec{k}_T - \vec{p}_{T,12}\Rb^2\,\Lb\vec{k}_T - \vec{p}_{T,1}\Rb^2  }{p^2_{T2}}\,+\,
  \frac{k^2_T\,\Lb\vec{k}_T - \vec{p}_{T,1} - \vec{p}_{T,12}\Rb^2  }{p^1_{T2}}\,-\,p^2_{T,12}\,-\,p^2_{T,12}  \frac{k^2_T\,\Lb\vec{k}_T- \vec{p}_{T,12}\Rb^2  }{p^1_{T2} p^2_{T,2}}\Bigg\}  \Bigg)\nn\\
&\times&\,\Bigg(\int d^2 l_{T}\,\,\phi^{N}_G\Lb Y - y_1; \vec{l}_T, -\vec{l}_T -\vec{p}_{T,12}\Rb\,\,  \frac{1}{l^2_T\,\Lb\vec{l}_T+ \vec{p}_{T,12}\Rb^2 \Lb\vec{l}_T - \vec{p}_{T,2}\Rb^4} \phi^{N}_G\Lb y_2; \vec{l}_T- \vec{p}_{T2},-\vec{l}_T+ \vec{p}_{T2} \Rb\nn\\
&\times&
\Bigg\{\frac{\Lb\vec{l}_T + \vec{p}_{T,12}\Rb^2\,\Lb\vec{l}_T - \vec{p}_{T,1}\Rb^2  }{p^2_{T2}}\,+\,
  \frac{l^2_T\,\Lb\vec{l}_T - \vec{p}_{T,2} \Rb^2  }{p^1_{T2}}\,-\,p^2_{T,12}\,-\,p^2_{T,12}  \frac{l^2_T\,\Lb\vec{l}_T+ \vec{p}_{T,12}\Rb^2  }{p^1_{T2} p^2_{T,2}}\Bigg\} 
\Bigg)\eea 
 The products  of $G_\mu G_\mu$  are written in \eq{RN4}
  $\{ \dots\}$ explicitly using \eq{A2}, and $\phi^N_D $ are
 the solution of \eq{RD2}.  Recall that
 $\phi^N_D\Lb \vec{k}_T,-\vec{k}_T + \vec{Q}_T\Rb$ vanishes
 both at $k_T \to 0$ and $\vec{k}_T - \vec{Q}_T \to 0$. Since
 the products of $G_\mu$ vanish at $\vec{k}_T - \vec{p}_{T1} \,\to\,0$
 or $\vec{l}_T - \vec{p}_{T2} \,\to\,0$, respectively, we can conclude
 that the integrals over $k_T$ and $l_T$ do not have large contributions
 at $k_T$ of the order of $p_{T1}$, and at $l_T$ of the order of $p_{T2}$. 
 
   %%%%%%%%%%%%%%%%%%%%%%%%%%%%%%%%%%%%%%%%%%%%%%%%%%%%%%
 %%%%%%%%%%%%%%%%%%%%%%%%%%%%%%%%%%%%%%%%%%%%%%%%
In the appendix we show that  the typical value of $Q_T$ for
 the BFKL Pomeron  $\phi^N_G\Lb Y; k'_T, k_T,Q_T\Rb$ is determined
 by the  smallest  value of transverse momentum $ Q_T \sim 
min\{k;_T,k_T\}$.
 In our case  this means that $Q_T =  p_{T,12}\, \sim 
1/R_N\,\gg\,p_{T1}$
 and/or $p_{T2}$.

Therefore, we can re-write \eq{RN4} as follows:
\bea \label{RN5}
\frac{d^2 \sigma}{d y_1 \,d y_2  d^2 p_{T1} d^2 p_{T2}}\Lb \fig{pdipom}\Rb\,&=&\,\,\h \Lb\frac{2 \pi \bas}{C_F}\Rb^2 \frac{1}{p^2_{T1}\,p^2_{T2}}\, \int d^2 Q_T  N\Lb p_{T,12}\Rb\,G_D\Lb \vec{Q}_T -  \vec{p}_{T,12}\Rb \nn\\
&\times&\,\Bigg(\int d^2 k_{T}\,\,\phi^{N}_G\Lb Y - y_1; \vec{k}_T, -\vec{k}_T \Rb\,  \phi^{N}_G\Lb y_2; \vec{k}_T- \vec{p}_{T1},-\vec{k}_T+ \vec{p}_{T1} \Rb\Bigg)\nn\\
&\times&\Bigg(\int d^2 l_{T}\,\,\phi^{N}_G\Lb Y - y_1; \vec{l}_T, -\vec{l}_T \Rb\, \phi^{N}_G\Lb y_2; \vec{l}_T- \vec{p}_{T2},-\vec{l}_T+ \vec{p}_{T2} \Rb\Bigg)
\eea 
In \eq{RN5} we introduce $\phi^N_G\Lb \vec{k}_t,-\vec{k}_T\Rb = 
(1/k^2_T)\phi^N_G\Lb \vec{k}_t,-\vec{k}_T\Rb(\eq{RN4})$. Comparing
 \eq{RN2} and \eq{RN5}, one can see that the correlation function is
 equal to

\beq \label{RN6}
C\Lb R^2_N\,p^2_{T,12}\Rb\,\,=\,\,N\Lb p^2_{12}\Rb
\eeq
for $\bas(y_1 - y_2) \leq 1$.

The three terms of $N\Lb Q_T\Rb$ are shown in \fig{n}-a.
 The first contribution $N\Lb Q_T\Rb\,=\,g^2\Lb Q_T\Rb$, can
  easily be  evaluated from the differential  elastic cross section,
 which is proportional to $g^4\Lb Q_T \Rb$. Recall, that the BFKL
 Pomeron does not generate the shrinkage of the diffraction peak
  seen in  the experimental data. This indicates that the
 exchange of the single BFKL Pomeron is not  sufficient to describe 
the
 high energy amplitude, and we need to use a more phenomenological 
approach
 to describe the elastic contribution to the correlation function (see
 Ref.\cite{GLMBE} in which we tried to describe these correlations
 using a particular model for high energy scattering, which is based
 on CGC/saturation approach).

For the onium, $g\Lb Q_T\Rb$ can be calculated (see \fig{n}-b and \eq{B5}) 
in the following way
\bea \label{RN7}
g\Lb Q_T\Rb\,=\,V^{\rm onium}\Lb  \vec{Q}_T\Rb\,&=&\,\int d^2 k'_T I_P\Lb \vec{k}_T, - \vec{k'}_T + \vec{Q}_T\Rb V^{\rm pr}\Lb  \vec{k'}_T, \vec{Q}_T\Rb\,\nn\\
&=&\,\int d^2 k'_T\Lb F\Lb Q_T\Rb - F\Lb 2 \vec{k'}_T - \vec{Q}_T\Rb\Rb\, V^{\rm pr}_{\h}\Lb \vec{k'_T},\vec{Q}_T\Rb
\eea
where $V^{\rm pr}$ is determined by \eq{B4}.
In \eq{RN7} $k'_T \sim 1/R_N \,\ll\,k_T$. 
Assuming that $F\Lb Q_T\Rb$  of \eq{IF1}  is equal to
 $1/\Lb 1 + R^2_N \,Q^2_T\Rb$,  we  find that at large $Q_T$, 
$g\Lb R_N \,Q_T\Rb$  decreases as $1/Q_T$.

%%%%%%%%%%%%%%%%%%%%%%%%%%%%%%%%%%%%%%%%%%%%%%%%

The second term of \fig{n}-b can be  evaluated from the process of
 diffraction dissociation in the region of small masses. However,
 we need to use a model for this term to be able to extract its $Q_T$ 
dependence
 from the experimental data. For example, we can replace the sum of
 possible produced diffractive states by one state, as has been done in
 Ref.\cite{GLMBE}. For the onium  state this term has the following form
\beq \label{RN8}
N_{\rm diff}\Lb Q_T\Rb\,=\,\int d^2 k'_T \int d^2 l'_T \, I_P\Lb
 \vec{k}_T,\vec{l}_T, - \vec{l}_T + \vec{Q}_T,-\vec{k}_T+  \vec{Q}_T\Rb 
\, V^{\rm pr}\Lb  \vec{k'}_T; \vec{Q}_T\Rb\,\, V^{\rm pr}\Lb   \vec{l'}_T;
 \vec{Q}_T\Rb
\eeq
where $I_P$ is taken from \eq{IF2}.

Using \eq{B4} we calculate $N_{\rm diff}\Lb Q_T\Rb$ 
 which decreases as $1/Q^2_T$ at large $Q_T$.
 
 ~

\begin{figure}[h]
 \includegraphics[width=11cm]{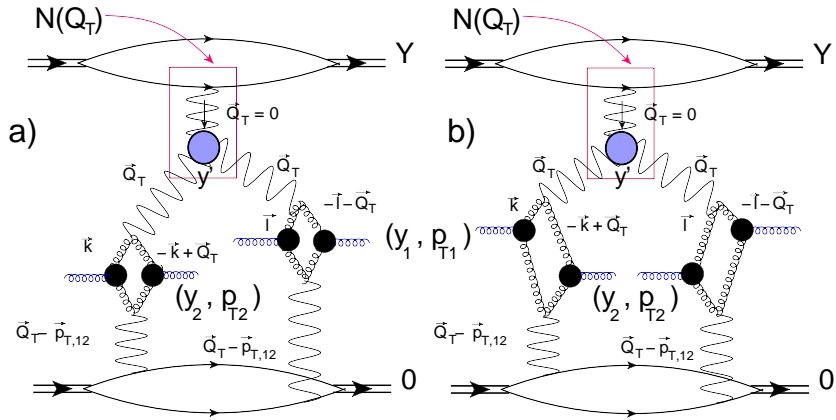}
   \protect\caption{ The large mass diffraction contribution to
 $N\Lb Q_T\Rb$: the source for the correlation length of about
 $1/Q_s$. The blue blob denotes the triple BFKL Pomeron vertex.
 The red  square indicates the contribution of $N\Lb Q_T\Rb$, which
 includes the integration over rapidity $y'$.}
\label{lmdiff}
   \end{figure}

  %%%%%%%%%%%%%%%%%%%%%%%%%%%%%%%%%%%%%%%%%%%%%%%
{\boldmath{\subsection{ $ R_c \,\,\propto\,\, 1/Q_s$}}}
%%%%%%%%%%%%%%%%%%%%%%%%%%%%%%%%%%%%%%%%%%%%%
The last term in \fig{n}-a, gives the contribution of large mass
 production in the diffraction dissociation process. The $Q_T$
 dependence of this term, is determined by the triple BFKL Pomeron
 vertex in perturbative QCD (see \fig{lmdiff}). Therefore,  this
 term generates correlations,  whose  length is determined by the BFKL
 Pomeron structure, and it  is closely related to the typical saturation
 momentum $Q_s$.

   %%%%%%%%%%%%%%%%%%%%%%%%%%%%%%%%%%%%%%%%%%%%%%%%%%%%%%
\begin{figure}[h]
 \includegraphics[width=11cm]{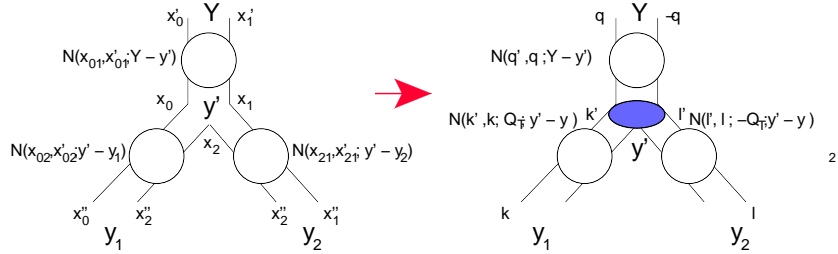}
   \protect\caption{ The triple BFKL Pomeron vertex in coordinate and 
momentum representations. The blue blob denotes the triple Pomeron vertex}
\label{3p}
   \end{figure}

 %%%%%%%%%%%%%%%%%%%%%%%%%%%%%%%%%%%%%%%%%%%%%%%%
 Comparing \fig{lmdiff} with \fig{pdspom} and \fig{pdipom}, one can see
 that the difference is only in
 expression for $N\Lb Q_T\Rb $ which has the following form
 \beq \label{RQ1}
 N_{\rm large \,mass\,diffraction}\Lb Q_T\Rb\,\,=\,\,\int \,d\,y'\, \phi^N_G\Lb Y - y'; Q_T=0; q_T, q'_T\Rb d^2 q'_T
 G_{3 \pom}\Lb q'_T; k'_T,  l'_T, Q_T\Rb
 \eeq
 We can obtain the form of $G_{3\pom}$ in  momentum space starting from
 the coordinate representation, where the contribution of the triple
 Pomeron diagram of \fig{3p} is known\cite{BRAUN,KLP}:
 \beq \label{3Pco}
 \bas \int \frac{d^2 x_0\,d^2 x_1\,d^2 x_2}{x^2_{01}\,x^2_{02}\,x^2_{21}} N\Lb x'_{01}, x_{01}; \vec{b} - \vec{b'}; Y-y'\Rb \,N\Lb x'_{02}, x_{02}; \vec{b'} - \h \vec{x}_{21}; y' - y_1\Rb \,N\Lb x'_{21}, x_{21}; \vec{b'} - \h \vec{x}_{02}; y' - y_2\Rb
 \eeq
 Introducing\cite{KOLEB}
 \beq \label{MOMSP}
 N\Lb x_{01},b; Y\Rb\,\,=\,\,x^2_{01}\int d^2 k d^2 Q_T \,e^{ i \vec{k}_T\cdot \vec{x}_{01}\,+\,i\,\vec{Q}_T\cdot \vec{b}}  N\Lb k_T, Q_T\Rb
 \eeq
 we see that \eq{3Pco} can be re-written in the form
 \beq \label{3Pmo}
 \bas \int d^2 q'_T \, N\Lb q_T,q'_T,Q_T=0, Y-y'\Rb\, G_{3 \pom}\Lb q'_T; k'_T,  l'_T, Q_T\Rb
   N\Lb k'_T,k_T,Q_T, y'-y_1\Rb  \, N\Lb l'_T,l_T,-Q_T, y'-y_2
   \Rb
 \eeq 
 
 with
 
 \beq \label{3P}
 G_{3 \pom}\Lb q'_T; k'_T,  l'_T, Q_T\Rb\,\,=\,\,\delta^{(2)}\Lb \vec{k'}_T \,-\vec{q'}\, + \h \vec{Q}_T\Rb\, \delta^{(2)}\Lb \vec{l'}_T \,-\vec{q'}\,- \h \vec{Q}_T\Rb\, 
 \eeq

     %%%%%%%%%%%%%%%%%%%%%%%%%%%%%%%%%%%%%%%%%%%%%%%%%%%%%%
 %%%%%%%%%%%%%%%%%%%%%%%%%%%%%%%%%%%%%%%%%%%%%%%% 
 
 In \eq{RQ1} we use the following notation for $\phi^N_G\Lb Y - y'; Q_T=0;
 k_i, k_f\Rb$: $Y-y'$ is the rapidity, $Q_T$ is the momentum transfer
   of the BFKL Pomeron, $k_i$ and $k_f$ are initial and final 
transverse
 momenta. Plugging  \eq{3P} in the general expression for the interference
 diagram of \fig{lmdiff}-b, we see that instead of \eq{RN4} we obtain
  \bea \label{QS1}
& &\frac{d^2 \sigma}{d y_1 \,d y_2  d^2 p_{T1} d^2 p_{T2}}\Lb \fig{lmdiff}-b\Rb\,\,=\nn\\
& &\,\,\h \Lb\frac{2 \pi \bas}{C_F}\Rb^2  \int d^2 Q_T  G_D\Lb \vec{Q}_T -  \vec{p}_{T,12}\Rb  \,\,\int \,d\,y'\,\int \phi^N_G\Lb Y - y'; Q_T=0; q_T, q'_T\Rb\ d^2 q'_T\nn\\
&\times&
\Bigg(G^{\rm BFKL}\Lb y' - y_1; p_{T,12}; \vec{q'}_T - \h \vec{p}_{T,12},k_T \Rb\, \, \frac{1}{k^2_T\,\Lb\vec{k}_T- \vec{p}_{T,12}\Rb^2 \Lb\vec{k}_T - \vec{p}_{T,1}\Rb^4} G^{\rm BFKL}\Lb y_2; 0 ;  \vec{k}_T- \vec{p}_{T1},k_N\Rb\nn\\
&\times&
  \Bigg\{\frac{\Lb\vec{k}_T - \vec{p}_{T,12}\Rb^2\,\Lb\vec{k}_T - \vec{p}_{T,1}\Rb^2  }{p^2_{T2}}\,+\,
  \frac{k^2_T\,\Lb\vec{k}_T - \vec{p}_{T,1} - \vec{p}_{T,12}\Rb^2  }{p^1_{T2}}\,-\,p^2_{T,12}\,-\,p^2_{T,12}  \frac{k^2_T\,\Lb\vec{k}_T- \vec{p}_{T,12}\Rb^2  }{p^1_{T2} p^2_{T,2}}\Bigg\}  \Bigg)\nn\\
&\times&\,\Bigg(\int d^2 l_{T}\,G^{\rm BFKL}\Lb y' - y_1; p_{T,12}, \vec{q'}_T + \h \vec{p}_{T,12}, l_T\Rb\,\,  \frac{1}{l^2_T\,\Lb\vec{l}_T+ \vec{p}_{T,12}\Rb^2 \Lb\vec{l}_T - \vec{p}_{T,2}\Rb^4} G^{\rm BFKL}\Lb y_2; 0;  \vec{l}_T- \vec{p}_{T2}, l_N \Rb\nn\\
&\times&
\Bigg\{\frac{\Lb\vec{l}_T + \vec{p}_{T,12}\Rb^2\,\Lb\vec{l}_T - \vec{p}_{T,1}\Rb^2  }{p^2_{T2}}\,+\,
  \frac{l^2_T\,\Lb\vec{l}_T - \vec{p}_{T,2} \Rb^2  }{p^1_{T2}}\,-\,p^2_{T,12}\,-\,p^2_{T,12}  \frac{l^2_T\,\Lb\vec{l}_T+ \vec{p}_{T,12}\Rb^2  }{p^1_{T2} p^2_{T,2}}\Bigg\} 
\Bigg)\eea  
The main difference  between \eq{QS1} and \eq{RN4}, is that $q'_T $ 
is larger
 than $q_T \approx\,1/R_N$. Indeed, the typical value of $q'_T = Q_s\Lb
 Y - y'\Rb\,\sim \Lb 1/R^2_N\Rb\exp\Lb \lambda \Lb Y - y'\Rb\Rb$, where
 $\lambda = \omega\Lb \gamma_{cr},0\Rb/(1 - \gamma_{cr}$ with
 $\gamma_{cr}=0.37$ in leading order of perturbative QCD \cite{KOLEB},

From \eq{PHIMY} one can see that each $\phi^N_D ( Y - y')
  \propto e^{\omega\Lb \h,0\Rb\,\Lb Y -  y'\Rb}$, $\phi^N_D
 ( y' -  y_1)\, \propto e^{\omega\Lb \h,0\Rb\,\Lb  y' - y_1\Rb}$,
 and $\phi^N_D ( y' - y_2) \propto e^{\omega\Lb \h,0\Rb\,\Lb 
 y' - y_2\Rb}$, since $\gamma  = \h + i \nu$ with small $\nu$.
 Therefore, integration over $y'$ results in $Y - y' \sim 1/\omega\Lb
 \h,0\Rb  \propto  1/\bas$ while $ y' - y_1$ and $y' - y_2$ are large
 (of the order of $Y$).  Since $Y - y' \ll y' - y_1 $($Y - y' \ll y'
 - y_2$) we can use the factorized formula of \eq{AF} for $\phi^{N}_G\Lb
 y' - y_1; p_{T,12}; \vec{q'}_T - \h \vec{p}_{T,12},k_T \Rb$ and for
 $\phi^{N}_G\Lb y' - y_1; p_{T,12}, \vec{q'}_T + \h \vec{p}_{T,12},
 l_T\Rb$. Using \eq{AF} we find that $p_{T,12}$ will be determined
 by the lowest momenta in the BFKL Pomeron with $y' - y_1$, and it will
 have the form

\beq \label{QS2}
C\Lb p_{T,12}\Rb\,\,\propto\,\,\int d^2 q'_T I_{-\gamma}\Lb q'_T\Rb\,V_\gamma\Lb \vec{q'}_T - \h \vec{Q}_T, \vec{Q}_T\Rb\,V_\gamma\Lb \vec{q'}_T + \h \vec{Q}_T, \vec{Q}_T\Rb
\eeq
where $V$ is determined by \eq{INTM},  and $Q_T = p_{T,12}$.
 In \eq{QS2} we can put $\gamma\,=\,\h$,  assuming $y' - y_1$ is 
suffiently
 large, that we can neglect $\nu$.

 %%%%%%%%%%%%%%%%%%%%%%%%%%%%%%%%%%%%%%%%%%%
\section{Bose-Einstein correlation function in the nucleon-nucleon 
interaction}
%%%%%%%%%%%%%%%%%%%%%%%%%%%%%%%%%%%%%%%

\begin{figure}[ht]
 \includegraphics[width=16cm]{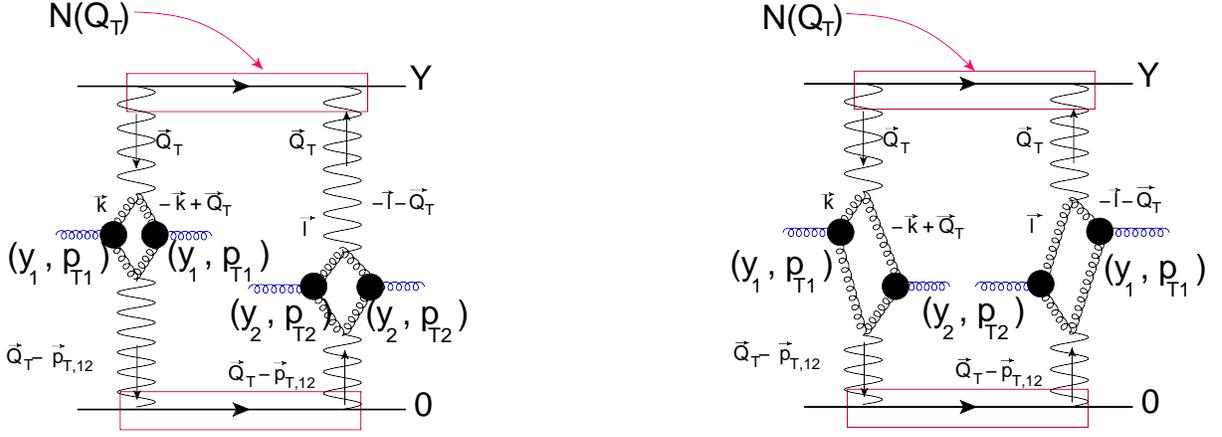}
   \protect\caption{ The Mueller diagram for the double inclusive
 production of two gluons with rapidities $y_1$ and $y_2$ and transverse
 momenta $\vec{p}_{T1}$ and $\vec{p}_{T2}$ 
   in  the nucleon-nucleon interaction. The BFKL Pomerons
    are denoted by wavy lines. The first diagram corresponds to the 
square
 of the amplitude for two parton showers production, while the second 
diagram
 describes the interference.
    }
\label{pppom}
   \end{figure}

In this section we discuss the Bose-Einstein correlations in nucleon-nucleon
 scattering. The Mueller diagrams for the square of the diagrams \fig{ddagk}-a
 and \fig{ddagk}-b, and for the interference diagrams, are shown in 
\fig{pppom}.
 This differs from the diagrams that have been discussed above,  
only in
 the appearance of the second $N\Lb Q_T\Rb$, which reflects the fact that 
we
 do not have small ( about $1/R_D$) momenta in this process.  Note, 
we can
 use perturbative QCD only if $p_{T1} \sim p_{T2} \gg 1/R_N$. Recalling that
 the $Q_T$ dependence of the BFKL Pomeron is determined by the smallest
 transverse momentum, we conclude that in \fig{pppom} the $Q_T$ dependence
 is determined by the function $N\Lb Q_T\Rb$. For the first two contributions
 to $N\Lb Q_T\Rb$ (see \fig{n}-a), this   is accurate to the order
of $1/\Lb
 R_N\,p_{T1}\Rb$. For the third contribution of the large mass 
diffraction,
 the  accuracy   is about $Q_s/p_{T1}$, where $Q_s $  denotes
the 
saturation momentum
 of the BFKL Pomeron with rapidity $Y - y'$.

 In spite of the fact that we indicate in \fig{n}-a the sources of
 experimental  information on  each contribution, the situation turns
 out to be more complicated. As an example, we discuss the elastic
 contribution. This gives $N\Lb Q_T\Rb = g^2\Lb Q_T\Rb$, where $g\Lb 
Q_T\Rb$
 is the Pomeron-hadron vertex. At first sight we can extract this vertex
 directly from the experimental values of $d \sigma_{el}/dt$.  However,
 this is certainly not correct. Indeed, the BFKL Pomeron cannot explain
 the shrinkage of the diffraction peak which is seen experimentally,
 and which gives almost a half of the slope of the elastic cross section
 for the energy range $W = 40 \,-\,7000\,GeV$\cite{TOTEM}. In the only
 model\cite{GLM2CH} for the soft interaction at high energy that is
 based on the BFKL Pomeron and Colour Glass Condensate (CGC)
 approach\cite{BK,JIMWLK}, the effective shrinkage of the
 diffraction peak stems from strong shadowing corrections, which lead to
 an elastic amplitude that is different from  that for the exchange 
of the BFKL
 Pomeron. However, it turns out that the most essential shadowing
 corrections  originate from the BFKL Pomeron interaction of
 two scattering hadrons. Such corrections do not contribute to the
 inclusive cross sections, as well as to the correlation due to AGK
 cutting rules\cite{AGK}.
 
It turns out to be an even more complicated problem to extract from the
 experimental data, the diffraction contribution to $N\Lb Q_T\Rb$ in the 
region
 of small masses. The lack of a theory, as well as insufficient 
experimental
 data, especially of the  momentum transfer distribution of the 
diffractively
 produced state with fixed mass, lead to the necessity of modeling  this
 process. The two extreme cases of such a modeling 
illustrates the difficulties: in our model \cite{GLM2CH} the rich variety
 of the produced states were replaced by a  single state: and in the 
constituent
 quark model\cite{CQM} the small mass diffraction stems from production of
 the state of free three constituent quarks. In our model the typical slope
 for $g_{diff}\Lb Q_T\Rb \propto \exp\Lb - B \,Q^2_T\Rb$ turns out to
 be 1/4 from the elastic slope, while in the CQM the
 size of the constituent quark
 is very  small.

  %%%%%%%%%%%%%%%%%%%%%%%%%%%%%%%%%%%%%%%%%%%%%%%%%%%%%%
 %%%%%%%%%%%%%%%%%%%%%%%%%%%%%%%%%%%%%%%%%%%%%%%%

 Taking the above into consideration, the uncertainties in the large 
mass diffraction term
 look small, and for  the triple BFKL Pomeron vertex, both the value and
 transverse momenta dependence follow directly from the Balitsky-Kovchegov
 equation\cite{BK}.
Bearing this in mind, we can write the expression for the interference
 diagram of \fig{pppom},  for the large mass diffraction contribution
 ( see \fig{n}). As we have discussed in this  case 
 $Q_T \sim Q_s\Lb Y - y'\Rb\,\,\ll\,\, \min\Big\{ p_{T1}(p_{T2}),
 Q_s\Lb y' - y_1\Rb\Big\} $,  $| \vec{Q}_T - \vec{p}_{T,12}|
 \sim\,Q_s\Lb Y - y'\Rb\,\,\ll\,\, \min\Big\{ p_{T1}(p_{T2}),
 Q_s\Lb y' - y_1\Rb\Big\} $, and $k'_T \sim \min\Big\{ p_{T1}(p_{T2}),
 Q_s\Lb y' - y_1\Rb\Big\}$ ( $l'_T \sim \min\Big\{ p_{T1}(p_{T2}),
 Q_s\Lb y' - y_1\Rb\Big\}$). Hence, we can use the factorized form
 for $\phi^G_N$ given by \eq{AF} and \eq{PHIMY}. 
 
 Finally, the large mass contribution for the interference diagram
 takes the form

\bea \label{NN1}
&&\frac{d^2 \sigma}{d y_1 \,d y_2  d^2 p_{T1} d^2 p_{T2}}\Lb \fig{pppom}, \mbox{interference diagram}\Rb\,\,=\,\,\nn\\
&&\,\h \frac{1}{N^2_c - 1}\Lb\frac{2 \pi \bas}{C_F}\Rb^2 \frac{1}{p^2_{T1}\,p^2_{T2}}   \,\,\int \,d\,y'\,\int \phi^N_G\Lb Y - y'; Q_T=0; q_N, q'_T\Rb \nn \\
&& \int  d^2 q'_T\,d^2 k_T
G^{\rm BFKL}\Lb y' - y_1; Q_T; \vec{q'}_T - \h \vec{Q}_{T}, k_T \Rb\, \int d y''\, G^{\rm BFKL}\Lb y_2  - y''; \vec{Q}_T - \vec{p}_{T,12} ;  \vec{k}_T- \vec{p}_{T1}, \vec{m}'_T - \h\Lb \vec{Q}_T - \vec{p}_{T,12}\Rb\Rb\nn\\
&&\times\,\,\int d^2 l_{T}\,d^2 m'_T \,G^{\rm BFKL}\Lb y' - y_1; Q_{T}, \vec{q'}_T + \h \vec{Q}_{T}, l_T\Rb\,\, G^{\rm BFKL}\Lb y_2 - y''; \vec{Q}_T - \vec{p}_{T,12};  \vec{l}_T- \vec{p}_{T2}, \vec{m}'_T + \h\Lb \vec{Q}_T - \vec{p}_{T,12}\Rb \Rb\nn\\
&&\times\,\,
\phi^N_G\Lb y'' ; Q_T=0; m_N, m'_T\Rb
\eea  

In the diagram for the square of the amplitude we can put
 $\vec{p}_{T,12}=0$.  Thus, 
the correlation function with the correlation length of the order
 of $1/Q_s\Lb Y - y'\Rb$, takes the following form

\beq \label{NN2}
C\Lb p_{T,12}/Q_s\Rb\,\,=\,\,\h \frac{1}{N^2_c - 1}\,\frac{N}{D}
\eeq
where
\bea \label{NNN}
&&  N \,=\,  \,\,\int d^2 Q_T\,\int \,d\,y'\,\int \phi^N_G\Lb Y - y'; Q_T=0; q_N, q'_T\Rb \nn \\
& & \,\int  d^2 q'_T\,d^2 k_T
G^{\rm BFKL}\Lb y' - y_1; Q_T; \vec{q'}_T - \h \vec{Q}_{T}, k_T \Rb\, \int d y''\,G^{\rm BFKL}\Lb y_2  - y''; \vec{Q}_T - \vec{p}_{T,12} ;  \vec{k}_T- \vec{p}_{T1}, \vec{m}'_T - \h\Lb \vec{Q}_T - \vec{p}_{T,12}\Rb\Rb\nn\\
&\times&\int d^2 l_{T}\,d^2 m'_T \,G^{\rm BFKL}\Lb y' - y_1; Q_{T}, \vec{q'}_T + \h \vec{Q}_{T}, l_T\Rb\,\, G^{\rm BFKL}\Lb y_2 - y''; \vec{Q}_T - \vec{p}_{T,12};  \vec{l}_T- \vec{p}_{T2}, \vec{m}'_T + \h\Lb \vec{Q}_T - \vec{p}_{T,12}\Rb \Rb\nn\\
&\times&
\phi^N_G\Lb y'' ; Q_T=0; m_N, m'_T\Rb
\eea
and
\bea \label{NND}
&&  D \,=\,  \,\,\int d^2 Q_T\,\int \,d\,y'\,\int \phi^N_G\Lb Y - y'; Q_T=0; q_N, q'_T\Rb \nn \\
& & \,\int  d^2 q'_T\,d^2 k_T
\phi^{N}_G\Lb y' - y_1; Q_T; \vec{q'}_T - \h \vec{Q}_{T}, k_T \Rb\, \int d y''\, \phi^{N}_G\Lb y_2  - y''; \vec{Q}_T - \vec{p}_{T,12} ;  \vec{k}_T- \vec{p}_{T1}, \vec{m}'_T - \h \vec{Q}_T \Rb\nn\\
&\times&\int d^2 l_{T}\,d^2 m'_T \,\phi^{N}_G\Lb y' - y_1; Q_{T}, \vec{q'}_T + \h \vec{Q}_{T}, l_T\Rb\,\, \phi^{N}_G\Lb y_2 - y''; \vec{Q}_T ;  \vec{l}_T, \vec{m}'_T + \h \vec{Q}_T  \Rb\nn\\
&\times&
\phi^N_G\Lb y'' ; Q_T=0; m_N, m'_T\Rb
\eea
  
The rather long algebraic expression of \eq{NNN} and \eq{NND} can be simplified using \eq{A17}  and they take the following forms
\bea \label{NNNS}
N\,&=&\,\int d^2 Q_T\,\int d y' e^{ 2 \omega\Lb\h,0\Rb \,y'}  \,d^2 q'_T \phi^N_G\Lb Y - y'; Q_T=0; q_N, q'_T\Rb\frac{V_{\h }\Lb \vec{q'_T},\vec{Q}_T - \h \vec{p}_{T,12} \Rb\,V_{\h }\Lb \vec{q'_T},\vec{Q}_T + \h \vec{p}_{T,12} \Rb}{ |\vec{Q}_T - \h \vec{p}_{T,12} |\,|\vec{Q}_T + \h \vec{p}_{T,12}|}\nn\\
&\times& \,\,\int d y'' e^{ 2 \omega\Lb\h,0\Rb \,y''}  \,d^2 l'_T \phi^N_G\Lb  y''; Q_T=0; l_N, l'_T\Rb\frac{V_{\h }\Lb \vec{l'_T},\vec{Q}_T  \Rb\,V_{\h }\Lb \vec{l'_T},\vec{Q}_T \Rb}{Q^2_T} \label{NNNS}\\
D\,&=&\,\int d^2 Q_T\,\int d y' e^{ 2 \omega\Lb\h,0\Rb \,y'}  \,d^2 q'_T \phi^N_G\Lb Y - y'; Q_T=0; q_N, q'_T\Rb\frac{V_{\h }\Lb \vec{q'_T},\vec{Q}_T  \Rb\,V_{\h }\Lb \vec{q'_T},\vec{Q}_T  \Rb}{ Q^2_T}\nn\\
 &\times& \,\,\int d y'' e^{ 2 \omega\Lb\h,0\Rb \,y''}  \,d^2 l'_T \phi^N_G\Lb  y''; Q_T=0; l_N, l'_T\Rb\frac{V_{\h }\Lb \vec{l'_T},\vec{Q}_T  \Rb\,V_{\h }\Lb \vec{l'_T},\vec{Q}_T \Rb}{Q^2_T}  \label{NNDS}\eea

\begin{figure}[ht]
 \includegraphics[width=16cm]{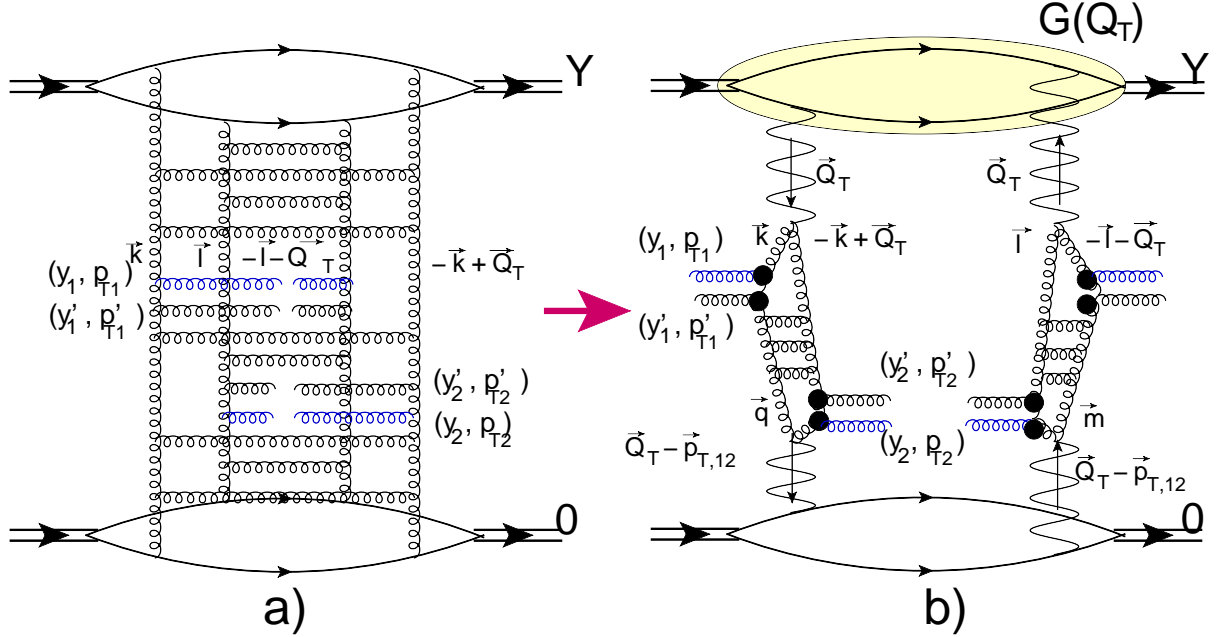}
   \protect\caption{ The double inclusive production of two gluons
 with rapidities $y_1$ and $y_2$ in the case of large  $| y_1 - y_2|$
 ( $ \bas\,| y_1 - y_2|\,\gg\,1$)  and transverse momenta $\vec{p}_{T1}$
 and $\vec{p}_{T2}$,  for the exchange of two BFKL Pomerons which are 
denoted
 by wavy lines. This diagram is the LLA generalization of 
\protect\fig{ddipom}.
 The  solid lines denote nucleons in the deuterons, which are 
 illustrated by
 double lines. Note, that the ladder, shown in \protect\fig{ddipomlay}-b, 
 represents
 the BFKL Pomeron with the momentum transferred $\vec{p}_{T1} + \vec{p}'_{T1}
  \approx \vec{p}_{T1}$ for $p'_{T1} \ll p_{T2}$ (see the text).  }
\label{ddipomlay}
   \end{figure}

 %%%%%%%%%%%%%%%%%%%%%%%%%%%%%%%%%%%%%%%%%%%
 {\boldmath
\section{$\bas \Lb y_1 \,-\,y_2\Rb\,\,\gg\,\,1$}}
%%%%%%%%%%%%%%%%%%%%%%%%%%%%%%%%%%%%%%%
All our previous estimates were performed for  small rapidity difference:
 $\bas | y_1 - y_2|\,\leq\,1$. In this section we  discuss 
 large rapidity differences ($ \bas | y_1 - y_2|\,\geq\,1$). For simplicity,
 we 
consider only correlations with the typical length of the order of $R_D$. In
 other words, we discuss the generalization of \fig{ddspom} and \fig{ddipom}
 to the case of large $y_{12} =  | y_1 - y_2|$. This generalization is shown
 in \fig{ddipomlay} for the interference diagrams.  The new features here 
are  that
 at rapidity $y'_1 < y_1$, we 
need to emit an additional gluon, and integrate over both its rapidity 
($y'_1$)
 and its transferred momentum ($p'_{T1}$).  Indeed, without this emission the
 ladder between rapidities $y'_1$ and $y'_2$ in \fig{ddipomlay}-b will be in
 the octet state of  color $SU_3$. The main idea is, that the principle 
contribution
 stems from $p'_{T1} \ll p_{T1}$. In this case the BFKL Pomeron between
 rapidities $y_1'$ and $y'_2$ has the momentum transfer which is equal
 to $p_{T1}$. After emission of two extra gluons with rapidities 
$y'_2$ and
 $y_2$, we obtain that the lower  BFKL Pomeron, that has momentum transfer 
$p_{T,12}$,
 as in \fig{ddipom}-b. In this diagram $Q_T \propto 1/R_D$, and can be put
 equal to zero in all parts of the diagrams, except of $G\Lb Q_T\Rb $ and
 $G\Lb \vec{Q}_T - \vec{p}_{T,12}\Rb$.

  %%%%%%%%%%%%%%%%%%%%%%%%%%%%%%%%%%%%%%%%%%%%%%%%%%%%%%
\begin{figure}[ht]
 \includegraphics[width=4cm]{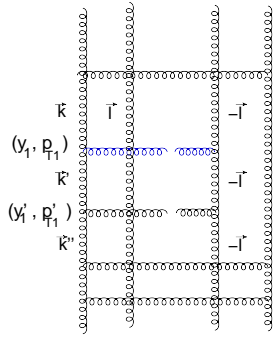}
   \protect\caption{ The part of the diagram of \fig{ddipomlay}-a
 with the vertex of emission of two gluons.  }
\label{ddnv}
   \end{figure}

 %%%%%%%%%%%%%%%%%%%%%%%%%%%%%%%%%%%%%%%%%%%%%%%%
First,  we need to integrate over $ p'_{T1}$. The vertex of the emission
 is shown in  \fig{ddnv} which can be written as

\beq \label{NV1}
\bas^2\,\Gamma_\mu\Lb \vec{k}_T,\vec{p}_{T1}\Rb\,\Gamma_\mu\Lb \vec{l}_T,\vec{p}_{T1}\Rb\,\frac{1}{k'^2_T\,l'^2_T}
\Gamma_\nu\Lb \vec{k}'_T,\vec{p}_{T1}\Rb\,\Gamma_\nu\Lb \vec{l}'_T,\vec{p}_{T1}\Rb\,\frac{1}{k''^2_T\,l''^2_T}\eeq
with $\vec{k}'_T =\vec{k}_T - \vec{p}_{T1}$ and $ \vec{k}''_T =\vec{k}'_T - \vec{p}'_{T1} = \vec{k}_T - \vec{p}_{T1} -  \vec{p}'_{T1}$.
Plugging in \eq{LIP},\eq{A2} and\eq{RD3} one can see that \eq{NV1} takes the form
\bea \label{NV2}
&&\bas^2\,\int d^2 p'_{T1} \frac{1}{p^2_{T1}}\Bigg\{ \frac{k^2_T}{k'^2_T}\,+\,\frac{l^2_T}{l'^2_T}\,-\,\frac{p^2_{T1}}{k'^2_T\,l'^2_T}\Bigg\}\,\frac{1}{p'^2_{T1}}\Bigg\{ \frac{k'^2_T}{k''^2_T}\,+\,\frac{l'^2_T}{l''^2_T}\,-\,\frac{p'^2_{T1}}{k''^2_T\,l''^2_T}\Bigg\}\,\\
&&\xrightarrow{p'_{T1} \ll k'_T ( l'_T) }\,2\,\bas^2\,\int^{k'_T}\frac{d^2 p'_{T1}}{p'^2_{T1}}\left[ \frac{1}{p^2_{T1}}\Bigg\{ \frac{k^2_T}{k'^2_T}\,+\,\frac{l^2_T}{l'^2_T}\,-\,\frac{p^2_{T1}}{k'^2_T\,l'^2_T}\Bigg\}\right]\,\,=\,\,
2\,\bas^2\,\int^{{\rm min} \{k'_T, l'_T\}}\frac{d^2 p'_{T1}}{p'^2_{T1}}\,K\Lb \vec{k}_T - \vec{l}_T,\vec{k}_T,\vec{k}'_T\Rb  \nn
\eea
Note that the term $\Big[\dots\Big]$ is the same as in \fig{ddipom} and
 $K\Lb \vec{k}_T - \vec{l}_T,\vec{k}_T,\vec{k}'_T\Rb$ is given by \eq{RD3}.

  %%%%%%%%%%%%%%%%%%%%%%%%%%%%%%%%%%%%%%%%%%%%%%%%%%%%%%
%%%%%%%%%%%%%%%%%%%%%%%%%%%%%%%%%%%%%%%%%%%%%%%%%%%%%%
 Finally,  we obtain the following expression for the interference diagram
 of \fig{ddnv}
 
 \bea \label{NV3}
&&\frac{d^2 \sigma}{d y_1 \,d y_2  d^2 p_{T1} d^2 p_{T2}}\Lb\fig {ddipomlay}-b \Rb\,\,=\,\,2 \Lb\frac{2 \pi \bas}{C_F}\Rb^2  \bas^2 \int d^2 Q_T  G_D\Lb Q_T\Rb\,G_D\Lb \vec{Q}_T -  \vec{p}_{T,12}\Rb\nn \\
&&\times\,\int d^2 k_{T}\,\,\int d^2 l_{T}\,\,\phi^{N}_G\Lb Y - y_1; \vec{k}_T, -\vec{k}_T \Rb\,
 \,\phi^{N}_G\Lb Y - y_1; \vec{l}_T, -\vec{l}_T\Rb\, \nn\\
 &&
 \int^{{\rm min} \{k'_T, l'_T\}}\int^{y_1} d y'_1 \,\frac{d^2 p'_{T1}}{p'^2_{T1}}\,K\Lb \vec{k}_T - \vec{l}_T,\vec{k}_T,\vec{k}'_T\Rb \int_{y_2} d y'_2 \,\int^{{\rm min} \{q'_T, m'_T\}} \frac{d^2 p'_{T2}}{p'^2_{T2}}\,K\Lb \vec{q}_T - \vec{m}_T,\vec{q}_T,\vec{q}'_T\Rb  \nn\\
 &&
 \int d^2 q_T \int d^2 m_T\, G^{\rm BFKL}\Lb y'_1- y'_2; \vec{p}_{T1}; \vec{k'}_T- \vec{p}'_{T1}, \vec{q}'_T + \vec{p}'_{T2}\Rb
  G^{\rm BFKL}\Lb y'_1- y'_2; - \vec{p}_{T1}; \vec{l}'_T - \vec{p}'_{T1} , \vec{m}'_T + \vec{p}'_{T2}\Rb\nn\\
 &&  \phi^{N}_G\Lb y_2; \vec{q}_T- \vec{p}_{T1},-\vec{q}_T+ \vec{p}_{T2} \Rb
\,\phi^{N}_G\Lb y_2; \vec{m}_T, -\vec{m}_T\Rb\eea

 In \eq{NV3} we put $Q_T = 0$ everywhere, except  in $ G_D\Lb Q_T\Rb$
 and $G_D\Lb \vec{Q}_T -  \vec{p}_{T,12}\Rb$, since $Q_T \,\sim\,1/R_D
 \ll $ all other momenta. At first sight \eq{NV3} gives the cross 
section which is suppressed as $\bas^2$ in comparison with \eq{RD5}.
 However, the integration over $y'_1$ and $y'_2$ leads to $1/\bas^2$
 contributions, resulting in a cross section of the order of $\bas^2$.
 One can also see, that the cross section  does not depend on the 
rapidity
 difference $y_{12}$ for the large values of this difference.
 
 The generalization to other cases,  which we have considered above, is
 straightforward, and we not discuss it here.

 %%%%%%%%%%%%%%%%%%%%%%%%%%%%%%%%%%%%%%%%%%%

\section{Conclusions}
%%%%%%%%%%%%%%%%%%%%%%%%%%%%%%%%%%%%%%% 
 \subsection{Comparison with other estimates in perturbative QCD} 
\begin{figure}[ht]
    \centering
    \begin{tabular}{ccc}
      \includegraphics[width=10cm]{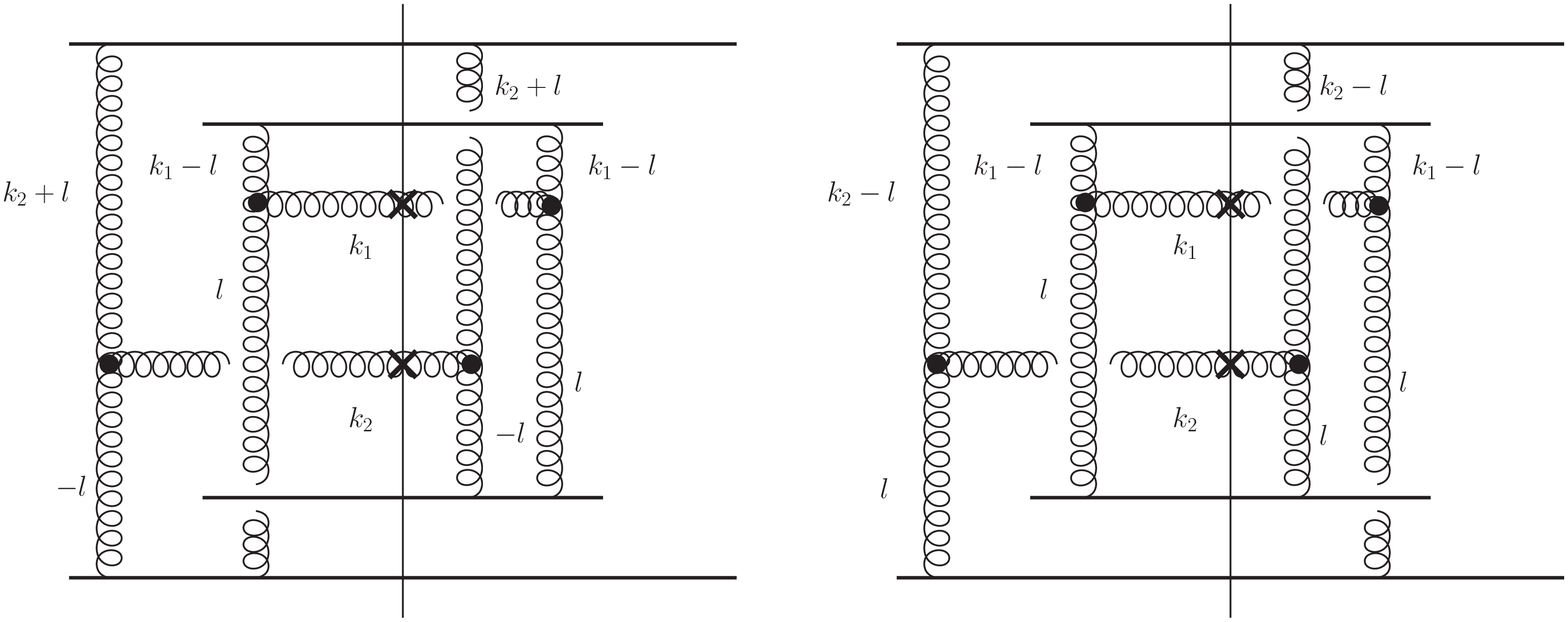} &~~~~~~ &\includegraphics[width=4.2cm]{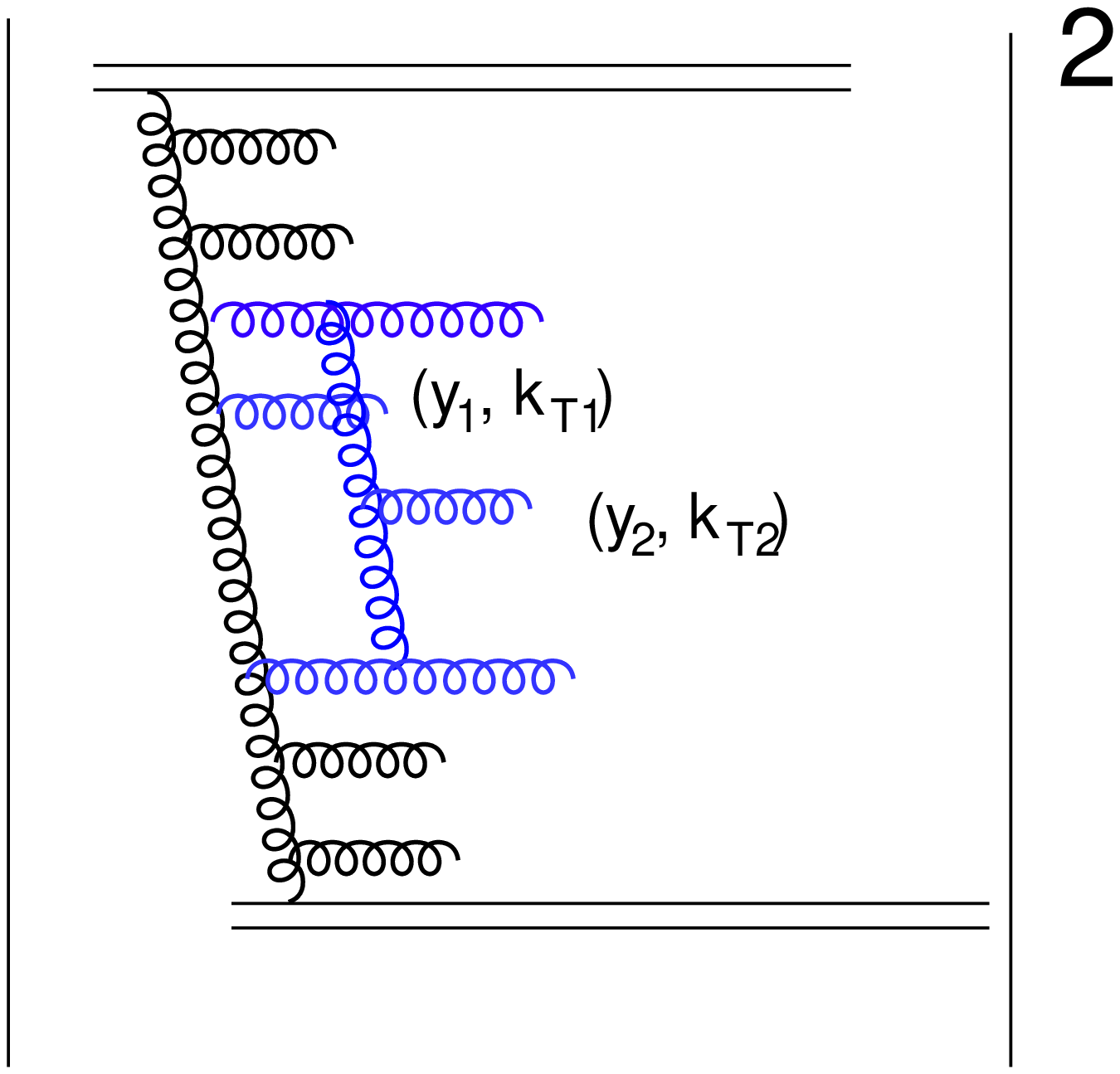}\\
     \fig{kov}-a & &\fig{kov}-b\\
      \end{tabular} 
     \protect\caption{\protect\fig{kov}-a is taken from Ref.\cite{KOWE}
 and describe the correlation in a one parton shower
 (see \protect\fig{kov}-b).}
      \label{kov}
      \end{figure}

 %%%%%%%%%%%%%%%%%%%%%%%%%%%%%%%%%%%%%%%
 The first estimate of the azimuthal correlations  due to the
 Bose-Einstein correlation in  perturbative QCD, was performed
 in Ref.\cite{KOWE}
 (see also Ref.\cite{KOLUCOR}). The diagrams, that were considered
 in these papers, are shown in \fig{kov}-a.  The observation is that
 these diagrams give rather strong azimuthal correlations, but they 
 are symmetric with respect to $\phi \to \pi  - \phi$, and only generate 
$v_n$
 with even $n$. The general origin of this symmetry  was discussed in
 section II-B for  slightly different diagrams.    In
 Refs.\cite{KOWE,KOLUCOR} the   $Q_T$ dependence was neglected leading to
 $\delta$-function contributions, which were smeared out by $Q_T$ 
dependance, 
 with $Q_T \sim 1/R$, where $R$ is the size of the interacting dipoles
 in \fig{kov}-a.
 
 %%%%%%%%%%%%%%%%%%%%%%%%%%%%%%%%%%%%%%%%%%%%%%%%  %%%%%%%%%%%%%%%%%
        
        %%%%%%%%%%%%%%%%%%%%%%%%%%%%%%%%%%%%%%%%%%%%%%%%%% 
  Since \fig{kov}-a describes  
  the production of two identical gluons in  the dipole-dipole 
amplitude, in the Born approximation of perturbative QCD,
  these diagrams are responsible for the azimuthal correlations in one
 parton cascade shown in \fig{kov}-b. It is worthwhile mentioning that 
 the  diagram of \fig{kov}-a leads to a contribution which is 
proportional
 to $\exp\Lb - \omega\Lb \h,0\Rb\,y_{12}\Rb$ and describes the 
correlations
 that decrease for large $y_{12}$.  Therefore, only for $\omega\Lb \h,0\Rb
 \,y_{12} \ll 1$,  can we consider this diagram as a source of 
correlations
 which are independent of $y_{12}$. 
 
  %%%%%%%%%%%%%%%%%%%%%%%%%%%%%%%%%%%%%%%%%%%%%%%%%%%%%%
\begin{figure}[ht]
 \includegraphics[width=9cm]{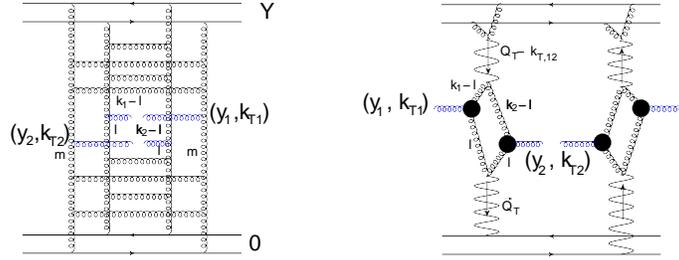} 
   \protect\caption{The  generalized  diagram of
\protect\fig{kov}-a, 
taking into account the gluon emission (two parton shower contribution).}
\label{kov1} 
   \end{figure}
%%%%%%%%%%%%%%%%%%%%%%%%%%%%%%%%%%%%%%%%%%%%%%%%%%%%%% 
Taking into account the emission of gluons, we can generalize the 
diagram of \fig{kov}-a to the
diagram of \fig{kov1}. We have 
considered this diagram above,  and have 
shown that there is no symmetry with respect of $\phi \to \pi - \phi$
 in such diagrams. Therefore, we conclude that the symmetry $\phi
 \to  \pi - \phi$, is a feature of  the azimuthal correlations in the one 
parton
 shower, in the Born approximation of perturbative QCD.

%%%%%%%%%%%%%%%%%%%%%%%%%%%%%%%%%%%%%%% 
 \subsection{Summary} 
 %%%%%%%%%%%%%%%%%%%%%%%%%%%%%%%%%%%%%%
In this paper, we found within the framework of perturbative QCD, that 
 the Bose-Einstein correlations due to two parton shower production, 
induce  azimuthal angle correlations, with three correlation lengths:
 the size of the deuteron, the proton radius, and the size of the BFKL
 Pomeron which is closely  related to the saturation momentum
 ($R_c \sim 1/Q_s$). These correlations are independent of the values
 of rapidities of produced gluons (long range rapidity correlations),
 and have no symmetry with respect to $\phi \to  \pi  - \phi$ ($ \vec{p}_{T1} 
 \to  - \vec{p}_{T1}$). Therefore,
 they give rise to $v_n$ for all values of $n$,  not only even values.

We reproduce the result of Refs.\cite{KOWE,KOLUCOR} which show this 
 symmetry in the Born approximation of perturbative QCD. However, even
 in the Born approximation, this symmetry depends  on the amplitude of
 the gluon - nucleon interaction at large distances, of about the
 nucleon size and, therefore, it  inherently has a non-perturbative
 nature. Replacing the nucleon by  an onium: the quark-antiquark 
bound
 state of heavy quarks, we see that symmetry $\phi \to \pi
 - \phi$ ($\vec{p}_{T1} \to - \vec{p}_{T1}$),  does not
 hold for distances of the order of the size of the onium.

We demonstrated that the azimuthal correlations with the correlation
 length ($R_c$)  of about the size of the deuteron, and the size of
 nucleon, stem from a non-perturbative contribution, and their estimates
 demand a lot of modeling due to the embryonic state of the theory in the
 non-perturbative region. However, the correlations with $R_c \sim 1/Q_s$
 have a perturbative origin, and can be 
evaluated in the framework of the Colour Gluon Condensate (CGC) approach.

We show that the two parton showers contributions, generate long range
 rapidity azimuthal angle correlations, which intuitively have been 
expected.  In other words, we demonstrate that the azimuthal
 angle correlations do not depend on $y_{12} = | y_1 - y_2|$
 for large values of $y_{12}$ ($\bas \,y_{12} \,\geq \,1$). 
 We illustrate that the correlation of Refs.\cite{KOWE,KOLUCOR}, actually 
describe
 the correlations in a one parton shower, and can be viewed, as 
independent 
of the
 rapidity difference, only in the narrow rapidity window $\bas y_{12}\ll 
1$.

  {\bf Acknowledgements} 
   We thank our colleagues at Tel Aviv University and UTFSM for
 encouraging discussions. Our special thanks go to    
Carlos Cantreras, Alex Kovner and Michel  Lublinsky for
 elucidating discussions on the
 subject of this paper. 
 
 This research was supported by the BSF grant   2012124, by   
 Proyecto Basal FB 0821(Chile) ,  Fondecyt (Chile) grant  
 1140842, and by   CONICYT grant PIA ACT1406.  
 ~
 
  \appendix

{\boldmath  \section{ $Q_T$ dependence of the BFKL Pomeron}}

%%%%%%%%%%%%%%%%%%%%%%%%%%%%%%%%%%%%%%%%

 The  impact parameter dependence of the BFKL Pomeron is well 
known\cite{LI},
 and it has the following form for the scattering of two dipoles ($r_1$ and
 $r_2$ ) at impact parameter $b$\cite{LI,NAPE}:
 \beq \label{NB}
N_\pom\Lb r_1,r_2; Y, b \Rb\,\,= \,\,
\int \frac{d \gamma}{2\,\pi\,i}
\,
e^{\omega(\gamma, 0 )\,Y}\,H^\ga\Lb w, w^*\Rb
\eeq 
 where
 \beq \label{OMEGA}
 \omega(\gamma, 0 )\,\,=\,\,\bas\Lb 2 \psi(1) - \psi(\gamma) - \psi(1 - \gamma)\Rb
 \eeq
 and where $\psi(z) $ is the Euler $\psi(z) = d \ln \Gamma\Lb z\Rb/d z $ 
(digamma function) (see Ref.\cite{RY} formula {\bf 8.360 - 8.367}).

  \bea 
 H^\ga\Lb w, w^*\Rb\,\,&\equiv&
\,\frac{ (\gamma - \h)^2}{( 
\gamma (1 - \gamma)
)^2} \Bigg\{b_\ga\,w^\gamma\,{w^*}^\gamma\,F\Lb\gamma, \gamma, 2\gamma, w\Rb\,
F\Lb\gamma, \gamma, 2\gamma, w^*\Rb
\,+ \nn\\
& &  b_{1 - \ga} w^{1 -
\gamma}{w^*}^{1-\gamma}
F\Lb 1 - \gamma, 1 -\gamma, 2 - 2\gamma, w\Rb\,F\Lb 1 - \gamma,1 -\gamma,2 
-2\gamma, w^*\Rb \Bigg\}\label{H}\\
 & \xrightarrow{w\,w^*\,\ll \,1}& \,\frac{ (\gamma - \h)^2}{( 
\gamma (1 - \gamma)
)^2} \Bigg\{b_\ga\,w^\gamma\,{w^*}^\gamma\,
\,+\,  b_{1 - \ga} w^{1 -
\gamma}{w^*}^{1-\gamma}\Bigg\}
\label{H1}
 \eea
where $F\,\equiv\,{}_2F_1$ is hypergeometric function \cite{RY}. In  \eq{H}
$w\,w^*$ is equal to
\beq \label{W}
w\,w^*\,\,=\,\,\frac{r^2_{1}\,r^2_{2}}{\Lb\vec{b} - \h\Lb\,\vec{r}_{1}\,
- \,\vec{r}_{2}\Rb\Rb^2
\,\Lb\vec{b} \,+\, \h \Lb\,\vec{r}_{1} \,- \,\vec{r}_{2}\Rb\Rb^2}
\eeq
and  $b_\ga$ is given by
\beq \label{BGA}
b_{\ga} \, = \, \pi^3 \, 2^{4(1/2 - \ga)} \, \frac{\Gamma \Lb\ga \Rb}{\Gamma \Lb 1/2 - \ga \Rb}
  \, \frac{\Gamma \Lb 1 - \ga  \Rb}{\Gamma \Lb 1/2 + \ga \Rb}.
\eeq

 From \eq{H} and \eq{W} we see that (i) $b$ is about of  the size   of 
the largest
 dipole ($ b \sim r_2 $ for $r_2 \gg r_1$);  and (ii) the scattering
 amplitude has a symmetry with respect to $\vec{b} \,\to\,- \vec{b}$.
 $Q_T$ is the conjugated variable to $b$, since
  \beq \label{NQ}
N_\pom \Lb r_1,r_2;  b; Y  \Rb\,\,= \,\,r^2_1 r^2_2 \int d^2 k\,d^2 k'
 e^{i \vec{k}_T\cdot\vec{r}_1\,+\,i \vec{k'}_T\cdot \vec{ r}_2}
\int  d^2 Q_T \,e^{i \vec{Q}_T \cdot \vec{b}}\,G^{\rm BFKL}\Lb Y; Q_T;
  k'_T, k_T \Rb\,
\eeq  
we see that the value of typical $Q_T \propto 1/r_2 \approx 1/R_N$.
 In \eq{NQ}   $G^{\rm BFKL}\Lb y - y'; Q_T;  k'_T, k_T \Rb$ denotes the
 BFKL Pomeron Green's function with the momentum transferred $Q_T$,
  and the transverse momenta of gluons $k_T$ at y and $k'_T$ at y'.
 The initial condition for the BFKL Green's function is the exchange
 of two gluons at $y = y'$.

In \eq{NQ}, the value of $r_1$ in our  problem, is about $1/p_{T1}$ or
 $1/p_{T2}$, and we  trust perturbative QCD calculations only
 if $p_{T1} \sim p_{T2} \,\gg\,1/R_N$.  Since $r_1 \ll r_2$ we can
 use \eq{H1} and take $w \,w^*$ to be equal to
\beq \label{WWSR}
w\,w^*\,\,=\,\,\frac{r^2_{1}\,r^2_{2}}{\Lb\vec{b} - \h\vec{r}_2\Rb^2
\,\Lb\vec{b} \,+\, \h \vec{r}_{2}\Rb^2}
\eeq

 Bearing in mind that
\beq \label{A1}
I_\gamma\Lb \vec{k}\Rb\,\,=\,\,\int \frac{d^2 r}{\Lb r^2\Rb^\gamma}\, e^{i \vec{k} \cdot\vec{r}}\,\,=\,\,2^{1 - 2 \gamma} \frac{\Gamma\Lb 1 - \gamma\Rb}{\Gamma\Lb \gamma\Rb} \frac{1}{\Lb k^2\Rb^{1 - \gamma}}
\eeq
Plugging \eq{A1} in \eq{NQ} we obtain
\bea \label{AN}
& &G^{\rm BFKL}\Lb \gamma;  \vec{Q}_T;  \vec{k'}_T,\vec{k}_T\Rb\,\,=\,\,\frac{\Lb\gamma -\h\Rb^2}{\Lb \gamma (1 - \gamma)\Rb^2}\nn\\
&\times &\Bigg\{b_\gamma\,
\int d^2 m'_T I_{1 - \gamma}\Lb \vec{k'}_T  - \vec{m'}_T\Rb\,I_\gamma\Lb \vec{m'} - \h \vec{Q}_T\Rb\,I_\gamma\Lb \vec{m'} + \h \vec{Q}_T\Rb\, I_{1 - \gamma}\Lb \vec{k}_T \Rb\,\nn\\
&+&\, b_{1-\gamma} \int d^2 m_T I_{\gamma}\Lb \vec{k'}_T  - \vec{m}_T\Rb\,I_{1 - \gamma}\Lb \vec{m} - \h \vec{Q}_T\Rb\,I_{1 -\gamma}\Lb \vec{m} + \h \vec{Q}_T\Rb\, I_{\gamma}\Lb \vec{k}_T \Rb\}
\Bigg\}\,
\eea

The integrals over $m'$ can be taken by replacing vector variables
 by the complex coordinates\cite{LI}
\beq \label{COMCO}
\vec{k}\,\,\to\,\,\rho_k\,=\,k_x \,+\,i\,k_y;~~~~~~\rho^*_k\,=\,k_x \,-\,i\,k_y;
\eeq
where $k_x$ and $k_y$ denote the $x$ and $y$ projections of $\vec{k}$.
 Using formula {\bf 3.197(1)} of Ref.\cite{RY} we can take integrals
 over $d^2 m' \,=\,d \rho_{m'}\,d \rho^*_{m'}$:
\bea
&&V_\gamma\Lb \vec{k'_T},\vec{Q}_T\Rb \,=\,\int d^2 m'_T I_{1 - \gamma}\Lb \vec{k'}_T  - \vec{m'}_T\Rb\,I_\gamma\Lb \vec{m'} - \h \vec{Q}_T\Rb\,I_\gamma\Lb \vec{m'} + \h \vec{Q}_T\Rb\,= \nn\\
&&\,\int d \rho_{m'}\,d \rho^*_{m'}I_{1 - \gamma}\Lb (\rho_k  - \rho_{m'}(\rho^*_k  - \rho^*_{m'})\Rb\,I_\gamma\Lb( \rho_{m'} + \h \rho_{Q}) ( \rho^*_{m'} + \h \rho^*_{Q})\Rb\, I_\gamma\Lb( \rho_{m'} - \h \rho_{Q}) ( \rho^*_{m'} - \h \rho^*_{Q})\Rb \nn\\
&& =\, 2^{2 - 4 \gamma}\Gamma^4\Lb 1 - \gamma\Rb\,\frac{1}{\Lb\Lb \vec{k}_T - \h \vec{Q}_T\Rb^2\Rb^\gamma}\,\frac{1}{\Lb Q^2_T\Rb^{1 - 2 \gamma}}\,F\Lb \gamma, \gamma,1,\frac{\rho_k + \h \rho_Q}{\rho_k - \h \rho_Q}\Rb \,\,F\Lb \gamma, \gamma,1,\frac{\rho^*_k + \h \rho^*_Q}{\rho^*_k - \h \rho^*_Q}\Rb \label{INTM}\\
&&= ~~\mbox{\Bigg({\bf 9.131(1)} of Ref.\cite{RY}\Bigg)}~~\,2^{2 - 4 \gamma}\Gamma^4\Lb 1 - \gamma\Rb \frac{1}{\Lb Q^2_T\Rb^\gamma}\,F\Lb \gamma,1 -  \gamma,1,\frac{\rho_k +  \rho_Q}{\rho_Q}\Rb\,F\Lb \gamma,1 -  \gamma,1,\frac{\rho^*_k +  \rho^*_Q}{\rho^*_Q}\Rb \label{INTM1} \\
&& =~~\mbox{\Bigg({\bf 9.132(2)} of Ref.\cite{RY}\Bigg)}~~\,\,2^{2 - 4 \gamma}\Gamma^4\Lb 1 - \gamma\Rb \frac{1}{\Lb \Lb \vec{k}_T + \vec{Q}_T\Rb^2\Rb^\gamma}\\
&& \,\times \Bigg\{ \frac{\Gamma\Lb 1 - 2 \gamma\Rb}{\Gamma^2\Lb 1 - \gamma\Rb}F\Lb \gamma, \gamma, 2 \gamma,\frac{\rho_Q}{\rho_k + \rho_Q}\Rb\,+\,  \Lb\frac{\rho_Q}{ \rho_k + \rho_Q}\Rb^{1 - 2 \gamma}\,\frac{\Gamma\Lb -1 + 2 \gamma\Rb}{\Gamma^2\Lb 1\gamma\Rb}F\Lb 1 - \gamma, 1 - \gamma, 2 (1 - \gamma),\frac{\rho_Q}{\rho_k + \rho_Q}\Rb\Bigg\}\nn  \label{INTM2}\\
&&\times \Bigg\{ \frac{\Gamma\Lb 1 - 2 \gamma\Rb}{\Gamma^2\Lb 1 - \gamma\Rb}F\Lb \gamma, \gamma, 2 \gamma,\frac{\rho^*_Q}{\rho^*_k + \rho^*_Q}\Rb\,+\,  \Lb\frac{\rho^*_Q}{\rho^*_k + \rho^*_Q}\Rb^{1 - 2 \gamma}\,\frac{\Gamma\Lb -1 + 2 \gamma\Rb}{\Gamma^2\Lb 1\gamma\Rb}F\Lb 1 - \gamma, 1 - \gamma, 2 (1 - \gamma),\frac{\rho*_Q}{\rho^*_k + \rho^*_Q}\Rb\Bigg\}\nn\eea

Plugging \eq{COMCO} in \eq{AN} one can see that  $\phi^N_G\Lb \gamma, \vec{k'}_T,\vec{k}_T, \vec{Q}_T\Rb$ can be written in the factorized form:
\beq \label{AF}
G^{\rm BFKL}\Lb \gamma; \vec{Q}_T; \vec{k'}_T,\vec{k}_T\Rb\,=\,V_\gamma\Lb \vec{k'_T},\vec{Q}_T\Rb\,I_\gamma\Lb \vec{k}_T\Rb\,\,+\,\,V_{1 - \gamma}\Lb \vec{k'_T},\vec{Q}_T\Rb\,I_{1 - \gamma}\Lb \vec{k}_T\Rb
\eeq

$\phi^N_G$ in the rapidity representation can be calculated as
\beq \label{PHIMY}
G^{\rm BFKL}\Lb Y; , \vec{Q}_T;  \vec{k'}_T,\vec{k}_T\Rb\,\,=\,\,\int^{\epsilon + i \infty}_{\epsilon - i \infty}\frac{d \gamma}{ 2 \pi i}\,e^{\omega\Lb \gamma, 0\Rb Y}\,G^{\rm BFKL}\Lb \gamma, \vec{k'}_T,\vec{k}_T, \vec{Q}_T\Rb
\eeq

Taking  the integral over $\gamma$ in \eq{PHIMY}  by the 
method of
 steepest descend \cite{KOLEB}, we  see that
for large $Y \gg 1$   the essential $\gamma = \h + i \nu$, where
 $\nu $ is small. Bearing this in mind, we can see from \eq{INTM1}
 that at large $Q_T \gg k'_T$, $G^{\rm BFKL}\Lb Y; , \vec{Q}_T; 
\vec{k'}_T,
\vec{k}_T\Rb\,\propto\,\,1/\Lb Q^2_T\Rb^\gamma \approx 1/Q_T$. At $Q_T \to
 0$ $Q_T \gg k'_T$ $G^{\rm BFKL}\Lb Y; , \vec{Q}_T; \vec{k'}_T,\vec{k}_T\Rb
\,\to\,$ Const.
Therefore, we conclude that the typical value $Q_T$ in the BFKL Pomeron,
 is about $k'_T$: the smallest transverse momenta.

At large $Y$ we can simplify \eq{PHIMY} using \eq{AF} and  the small 
size
 of $\nu$. Plugging  
$G^{\rm BFKL}\Lb Y; ,\vec{Q}_T; \vec{k'}_T,\vec{k}_T\Rb$ from \eq{AF}
in \eq{PHIMY},
 we have

\bea \label{A17}
G^{\rm BFKL}\Lb Y;  \vec{Q}_T; \vec{k'}_T,\vec{k}_T\Rb\,\,&=&\,\,\int^{\epsilon + i \infty}_{\epsilon - i \infty}\frac{d \gamma}{ 2 \pi i}\,e^{\omega\Lb \gamma, 0\Rb Y}\,2\,V_{\h + i\nu}\Lb \vec{k'_T},\vec{Q}_T\Rb\,I_\gamma\Lb \vec{k}_T\Rb\,\nn\\
&\xrightarrow{\nu \ll 1}&\,\,\frac{2\,C\Lb\gamma= \h\Rb}{k'_T\,k_T}V_{\h }\Lb \vec{k'_T},\vec{Q}_T\Rb\int^{+ \infty+ i \epsilon}_{-  \infty + i \epsilon} \frac{d \nu}{2 \pi}\,\,e^{\Lb \omega\Lb \h, 0\Rb \,- \,D\, \nu^2\Rb\, Y\,+ i \nu \ln\Lb k^2_T/k'^2_T\Rb}\nn\\
&=&\,\,\frac{2\,C\Lb\gamma= \h\Rb}{k'_T\,k_T}V_{\h }\Lb \vec{k'_T},\vec{Q}_T\Rb\Bigg(\sqrt{\frac{2 \pi}{D\,Y}}\,e^{  \omega\Lb \h, 0\Rb\,Y\,-\,\frac{\ln^2\Lb k'^2_T/k^2_T\Rb}{ 4\,D\,Y}}\Bigg)
\eea

Integral in \eq{A17} is taken in the saddle point approximation with
 $\nu_{SP}=i \ln\Lb k^2_T/k'^2_T\Rb/2\,D\,Y\,\,\ll\,\,1$ and $ 
 \omega\Lb \h= i \nu, 0\Rb\,\,=\,\,  \omega\Lb \h, 0\Rb\,\,-\,\,D\,\nu^2$.

~

 \section{ The BFKL Pomeron -  onium vertex}
 The scattering amplitude of a dipole of  size $x_{01}$ with an 
onium has the following form:
 \beq \label{B1}
 A\Lb Y, x_{01}; Q_T \Rb\,\,=\,\,\int d^2 x'_{01}\,e^{i \vec{x'}_{01} \cdot \vec{Q}_T}\,\Psi^*_{\rm onium}\Lb x'_{01}\Rb\,N\Lb Y; x'_{01},x_{01}, Q_T \Rb\, \Psi^*_{\rm onium}\Lb x'_{01}\Rb
 \eeq
 In the  momentum representation it  can be written as
 \beq \label{B2}
 A\Lb \gamma, k'_T,k_T,Q_T\Rb\,=\, \Lb F\Lb Q_T\Rb - F\Lb 2 \vec{k'}_T - \vec{Q}_T\Rb\Rb\int d^2 x'_{01} e^{- i \vec{k'}_T \cdot\vec{x'}_{01} -i \vec{k}_T \cdot\vec{x}_{01} } \,N\Lb \gamma; x'_{01},x_{01}, Q_T \Rb/x^2_{01}
 \eeq
  The amplitude $e^{- i \vec{k'}_T \cdot\vec{x'}_{01} -i \vec{k}_T \cdot\vec{x}_{01} } \,N\Lb \gamma; x'_{01},x_{01}, Q_T \Rb/x^2_{01}$ can be written in the factorized form of \eq{AF}:
  \beq \label{B3}
  e^{- i \vec{k'}_T \cdot\vec{x'}_{01} -i \vec{k}_T \cdot\vec{x}_{01} } \,N\Lb \gamma; x'_{01},x_{01}, Q_T \Rb/x^2_{01}\,\,=\,\,\,V^{\rm pr}_\gamma\Lb \vec{k'_T},\vec{Q}_T\Rb\,N_\gamma\Lb \vec{k}_T\Rb\,\,+\,\,V^{\rm pr}_{1 - \gamma}\Lb \vec{k'_T},\vec{Q}_T\Rb\,N_{1 - \gamma}\Lb \vec{k}_T\Rb
  \eeq
    where $V^{\rm pr}_\gamma\Lb \vec{k'_T},\vec{Q}_T\Rb$ is equal to
    \bea
       &&V^{\rm pr}_\gamma\Lb \vec{k'_T},\vec{Q}_T\Rb \,=\,\int d^2 m'_T I_{ - \gamma}\Lb \vec{k'}_T  - \vec{m'}_T\Rb\,I_\gamma\Lb \vec{m'} - \h \vec{Q}_T\Rb\,I_\gamma\Lb \vec{m'} + \h \vec{Q}_T\Rb\,= \label{B4} \\
&&\,\int d \rho_{m'}\,d \rho^*_{m'}I_{ - \gamma}\Lb (\rho_k  - \rho_{m'}(\rho^*_k  - \rho^*_{m'})\Rb\,I_\gamma\Lb( \rho_{m'} + \h \rho_{Q}) ( \rho^*_{m'} + \h \rho^*_{Q})\Rb\, I_\gamma\Lb( \rho_{m'} - \h \rho_{Q}) ( \rho^*_{m'} - \h \rho^*_{Q})\Rb \nn\\
&& =\, 2^{3 - 2\gamma}\frac{\Gamma\Lb 1 + \gamma\Rb\Gamma^2\Lb 1 - \gamma\Rb}{\Gamma\Lb- \gamma\Rb\,\Gamma^2\Lb \gamma\Rb}\,\frac{1}{\Lb\Lb \vec{k}_T - \h \vec{Q}_T\Rb^2\Rb^{1+\gamma}}\,\frac{1}{\Lb Q^2_T\Rb^{1 - 2 \gamma}}\,F\Lb1+ \gamma, \gamma,2,\frac{\rho_k + \h \rho_Q}{\rho_k - \h \rho_Q}\Rb \,\,F\Lb 1+\gamma, \gamma,2,\frac{\rho^*_k + \h \rho^*_Q}{\rho^*_k - \h \rho^*_Q}\Rb
\nn\eea
Finally, the BFKL Pomeron-onium vertex takes the form
\beq \label{B5}
V^{\rm onium}_\gamma\Lb Q_T\Rb\,=\,\int d^2  k'_T \Lb F\Lb Q_T\Rb - F\Lb 2 \vec{k'}_T - \vec{Q}_T\Rb\Rb\, V^{\rm pr}_\gamma\Lb \vec{k'_T},\vec{Q}_T\Rb
\eeq    
    
 \end{document}